\documentstyle[emulateapj]{article}
\def\etal{{et al.}\ }
\def\per{$^{-1}$}
\def\persq{$^{-2}$}
\def\hal{H$\alpha$}
\def\hbeta{H$\beta$}
\def\kms{km s$^{-1}$}

\def\msun{$M{\footnotesize{_{\odot}}}$}

\def\hst{\emph{HST}}
\def\lam{$\lambda$}
\def\ebv{$E(B-V)$}
\def\iras{\emph{IRAS}}
\def\sn{\emph{S/N}}

\begin{document}

\submitted{Accepted for publication in the Astrophysical Journal}

\title{Polarized Broad-Line Emission from Low-Luminosity Active Galactic
Nuclei}

\author{Aaron J. Barth}
\affil{Harvard-Smithsonian Center for Astrophysics, 60
Garden St., Cambridge MA 02138; abarth@cfa.harvard.edu}
\author{Alexei V. Filippenko and Edward C. Moran}
\affil{Department of Astronomy, University of California,
Berkeley CA 94720-3411; \\ alex@astro.berkeley.edu, emoran@astro.berkeley.edu}

\begin{abstract}

In order to determine whether unified models of active galactic nuclei
apply to low-luminosity objects, we have undertaken a
spectropolarimetric survey of of LINERs and Seyfert nuclei at the Keck
Observatory.  The 14 objects observed have a median \hal\ luminosity
of $8\times10^{39}$ erg s\per, well below the typical value of
$\sim10^{41}$ erg s\per\ for Markarian Seyfert nuclei.  Polarized
broad \hal\ emission is detected in three LINERs: NGC 315, NGC 1052,
and NGC 4261.  Each of these is an elliptical galaxy with a
double-sided radio jet, and the emission-line polarization in each
case is oriented roughly perpendicular to the jet axis, as expected
for the obscuring torus model.  NGC 4261 and NGC 315 are known to
contain dusty circumnuclear disks, which may be the outer extensions
of the obscuring tori.  The detection of polarized broad-line emission
suggests that these objects are nearby, low-luminosity analogs of
obscured quasars residing in narrow-line radio galaxies.  The nuclear
continuum of the low-luminosity Seyfert 1 galaxy NGC 4395 is polarized
at $p = 0.67\%$, possibly the result of an electron scattering region
near the nucleus.  Continuum polarization is detected in other
objects, with a median level of $p = 0.36\%$ over 5100-6100 \AA, but
in most cases this is likely to be the result of transmission through
foreground dust.  The lack of significant broad-line polarization in
most type 1 LINERs is consistent with the hypothesis that we view the
broad-line regions of these objects directly, rather than in scattered
light.

\end{abstract}


\section{Introduction}

Ever since the discovery of a hidden broad-line region (BLR) in NGC
1068 (\cite{am85}), spectropolarimetric observations have provided a
powerful means to detect fundamental similarities among the diverse
and often bewildering menagerie of active galactic nuclei (AGNs).
Unified models, based in large part on evidence from
spectropolarimetry, have been very successful at explaining the impact
of viewing angle on our classification of many different types of
high-luminosity AGNs (\cite{ant93}), but little is known about the
place of low-luminosity AGNs (LLAGNs) within the unification
framework.

The division between low- and high-luminosity AGNs is somewhat
arbitrary.  Ideally this distinction should be made on the basis of
bolometric luminosities, but these are not available for most nearby
AGNs, and there is no clear physical basis for choosing a particular
luminosity threshold for separating low- and high-luminosity objects.
One criterion which is adequate for most purposes and observationally
straightforward is the luminosity in the narrow \hal\ line.  Ho,
Filippenko, \& Sargent (1997a) note that typical Seyfert nuclei in the
Markarian catalog have \hal\ luminosities of $\sim10^{41}$ erg s\per\
(e.g., \cite{whi92}), while in the Ho \etal sample of 500 nearby
galaxies, 85\% of the AGNs have $L$(\hal) $< 10^{40}$ erg s\per.  Ho
\etal (1997a) propose that LLAGNs may be defined operationally as
those AGNs having $L$(\hal) $\leq 10^{40}$ erg s\per. We will adopt
this definition here, bearing in mind the ambiguities brought about by
anisotropic emission, the uncertain fraction of the bolometric
luminosity contributed by \hal, and the uncertainties in the distances
to these nearby galaxies.

Low-ionization nuclear emission-line regions, or LINERs (\cite{h80}),
are one class of LLAGNs to which unified models have not been applied
in the past.  About 15\% of LINERs have a broad component of the \hal\
emission line (\cite{hfsp97}), similar to the fraction of Seyferts
which are broad-lined, and we will refer to ``LINER 1'' and ``LINER
2'' nuclei as those in which broad \hal\ emission is detected or
undetected, respectively, in the total flux spectrum.  Despite some
observational effort in the past using 3-m and 4-m telescopes (see
\cite{ant93} and Barth 1998), it has never been determined whether the
faint broad \hal\ wings in LINER 1 nuclei are viewed directly or in
scattered light.  It is also not known whether some fraction of the
LINER 2 population may contain ``hidden'' type 1 nuclei.  The latter
question is of particular importance because some LINER 2 nuclei may
not be AGNs at all; it is sometimes possible to explain their optical
emission-line ratios and energetics in terms of shock heating
(\cite{ds95}) or ionization by hot stars (\cite{ft92}; \cite{shi92};
\cite{bin94}), rather than by AGN-like photoionization by a nonstellar
continuum source.  Since approximately one quarter of all nearby
galaxies have LINER 2 nuclei (\cite{hfs97demo}), our understanding of the
faint end of the AGN luminosity function depends crucially on
determining the power source in these objects.  The detection of
polarized nonstellar continua and broad emission lines in LINER 2
galaxies would provide convincing evidence that these objects do
contain genuine low-luminosity active nuclei.

Unfortunately, several difficulties conspire to make the detection of
polarization in LLAGNs extremely challenging.  The optical continua
are typically dominated by unpolarized starlight, and it is generally
impossible to detect any featureless continuum contribution to the
total flux spectrum.  Due to this starlight contamination, the
continuum polarization in LLAGNs is expected to be at a level of
$<1\%$, and such small intrinsic polarizations can easily be masked by
foreground interstellar polarization.  In some LLAGNs the central
engines may be hidden by dust lanes in the host galaxy (e.g.,
\cite{mr95}; \cite{mgt98}; \cite{bar98}), and obscuration of this sort
would not lead to any polarization signature, except possibly through
dust transmission.  In the well-studied case of NGC 1068, the nuclear
polarization is the result of scattering by electrons (\cite{mgm91};
\cite{ahm94}), but a similar electron-scattering medium might not be
present in low-luminosity, low-ionization AGNs.  Finally, some
fraction of LINER 2 nuclei may derive their power from hot stars,
rather than from an AGN, in which case no hidden BLR would be present
at all.  Therefore, it is reasonable to expect that very high
signal-to-noise ratios (\sn) will be required in order to detect
optical polarization in LLAGNs, and that most LLAGNs will not be
polarized at any detectable level other than from dust transmission.

Much progress has been made recently in searching for hidden BLRs in
starlight-dominated AGNs (e.g., \cite{wil95}; \cite{you96};
\cite{km98}; \cite{oli98}), but previous surveys have not concentrated
on LLAGNs or LINERs specifically.  In order to determine whether
polarized broad emission lines occur in LLAGNs, we have undertaken a
spectropolarimetric survey at the Keck Observatory.  Our first
detection of polarized broad-line emission, in the LINER NGC 1052, has
been published separately (Barth, Filippenko, \& Moran 1999a). This
paper presents the data we have collected to date for this project,
for a total of 14 objects.

\section{Observations}

The observations were made during four observing runs at the Keck-II
telescope, on 1997 December 20, 1998 January 17, 1998 March 7, and
1999 January 6 (UT).  Targets were selected from the LINERs and
Seyfert nuclei observed in the Ho \etal (1997a) spectroscopic
survey, and the well-known southern-hemisphere LINER NGC 1097 was
included as well.  The galaxies observed do not form a statistically
meaningful sample; our primary goal was to determine whether polarized
emission could be detected in the brightest LINERs, and we selected
objects having emission lines of high brightness and large equivalent
width.  Table \ref{journal} lists the galaxy sample.  Most of the
targets have faint broad \hal\ emisison in total flux (\cite{hfsp97}).
The narrow \hal\ luminosities listed in Table \ref{journal} are taken
from Ho \etal (1997c) or from other sources in the literature for
those objects which were not observed under photometric conditions by
Ho et al.  Five galaxies exceed the nominal $10^{40}$ erg s\per\
cutoff for classification as ``low-luminosity'' objects, but all have
luminosities below the typical $L$(\hal) of $10^{41}$ erg s\per\ for
Markarian Seyfert nuclei.  The median \hal\ luminosity of the sample
is $8 \times 10^{39}$ erg s\per.

Observations were obtained with the LRIS spectropolarimeter
(\cite{oke95}; \cite{coh97}), using a 600 grooves mm\per\ grating
blazed at 5000 \AA\ and a slit width of 1\arcsec.  The spectra covered
approximately 4320--6860 \AA, with a resolution of $\sim5-6$ \AA\ and
a pixel scale of 1.2 \AA\ bin\per.  Seeing and weather conditions were
$1\farcs3-1\farcs6$ with clear skies during the December 1997 run,
$1\farcs1-1\farcs3$ with thick clouds during the January 1998 run,
$0\farcs9-1\farcs1$ with clear skies during the March 1998 run, and
$1\farcs0-1\farcs1$ with thick clouds during the January 1999 run.
Exposure times for the individual targets are listed in Table
\ref{journal}.  Most exposures were obtained with the slit oriented at
the parallactic angle, but for a few objects at low airmass ($<1.3$)
the slit was oriented at P.A. = 90\arcdeg.  Total exposure times
ranged from 10 to 70 minutes per galaxy.  Due to the wide range in
emission-line and continuum fluxes of these galaxies, as well as to
rapid changes in atmospheric conditions during parts of these nights,
the resulting \sn\ varied from object to object.

The spectra were flat-fielded, extracted, and wavelength- and
flux-calibrated using IRAF.  Subsequent spectropolarimetric processing
was performed using routines written in IDL, following the methods
described by Miller, Robinson, \& Goodrich (1988) and Cohen \etal
(1997).  Uncertainties due to photon-counting statistics and detector
readout noise were propagated at every step of the reduction process,
yielding ``error spectra'' for the Stokes parameters $q$ and $u$.

Extractions were 2\farcs6--5\farcs2 wide, depending on the seeing and
the spatial profile of each individual galaxy.  The large extraction
widths were chosen in order to minimize the amplitude of spurious
features in the continuum; in narrower (1\arcsec-2\arcsec)
extractions, broad and narrow bumps appeared in the continuum
polarization spectra with amplitudes of a few tenths of a percent.
Such features in $p$ can arise from uncertainties in interpolating
counts in fractional pixels at the edges of the extraction aperture,
due to the finite sampling (0\farcs43 pixel size) of the galaxy's
spatial profile.  These spurious features in $p$ can be coherent over
a broad wavelength range, as they depend on the spatial profile of the
galaxy, the centering of the galaxy profile on the detector pixels,
and the accuracy of the spectral traces.  Since the amplitude of these
errors depends on the fraction of total flux contributed by the
outermost pixel on each side of the extraction aperture, extractions
were performed with apertures sufficiently wide that the outermost
pixels did not contribute substantially to the total flux.  This issue
only becomes problematic when one is searching for true features in
$p$ at the level of a few tenths of a percent in spatially extended
sources.  For point sources, such as the standard stars, the best
approach is to use a wide extraction together with an optimal
weighting algorithm (\cite{hor86}).  Consequently, the outermost
pixels in a wide aperture contribute negligibly to the total counts.

The extracted spectra were rebinned to 2 \AA\ bin\per\ prior to the
polarimetric analysis.  Cosmic-ray spikes were removed by comparing
adjacent pairs of extracted spectra.  To account for slight shifts in
wavelength among the eight individual extracted spectra from each
observing sequence, the spectra were aligned in wavelength to within
0.1 \AA\ by measuring their cross-correlations.  The instrumental
polarization response was checked each night using observations of
polarized and unpolarized standard stars, and the polarizations found
for these stars agreed well with previously published values
(\cite{tur90}).  The zeropoint of the polarization angle was
calibrated using the polarized standard stars as references.
Observations of unpolarized standard stars (Figure \ref{null}) showed
a well-behaved, flat response over the observed wavelength range and
an instrumental polarization of $<0.05\%$.

All measurements of emission-line and continuum polarization were
carried out on the $q$ and $u$ spectra, with the results converted to
$p$ and $\theta$ (the position angle of polarization) only as the
final step, following the methods outlined by Miller \etal (1988).
For display purposes, the rotated Stokes parameter was computed by
heavily smoothing the $\theta$ curve and rotating $q$ and $u$ by this
smoothed angle, to yield a single Stokes parameter containing
essentially all of the measured polarization.  As discussed by Miller
\etal (1988), the polarization $p$ has a peculiar error distribution
which becomes particularly problematic at low \sn, because the usual
formula $p = \sqrt{q^2 + u^2}$ is positive definite.  The rotated
Stokes parameter is equivalent to the degree of polarization $p$, but
its noise properties are better behaved.  The Stokes flux spectrum is
calculated as the product of total flux and the rotated Stokes
parameter, and gives the spectrum of the linearly polarized component
of the total flux.  The total flux, rotated Stokes parameter, and
Stokes flux spectra for each galaxy are shown in Figures
\ref{ngc315plot}-\ref{ngc4594plot} and the individual galaxies are
discussed in \S\ref{sectionindividual}.  We do not display the
$\theta$ spectra, as they are extremely noisy and convey little useful
information.

As a check on the error propagation, the Stokes parameter error
spectra can be compared with the actual pixel-to-pixel variations in
the Stokes parameter spectra.  In regions where the Stokes parameters
appear roughly constant, we generally find excellent agreement between
the propagated uncertanties in $q$ and $u$ and the standard deviations
of pixel values in the $q$ and $u$ spectra.  However, the quoted
uncertainties on $p$ reflect only the random errors due to
photon-counting statistics and readout noise.  At the very weak
polarization levels discussed here, systematic uncertainties in the
spectral extractions and calibrations can contribute appreciably to
the actual errors, causing faint broad features or undulations to
appear in $p$, as seen in some of our spectra.  Incompletely removed
cosmic-ray hits can also lead to spikes appearing in $p$.

\section{Measurements and Results}

The continuum polarization and $\theta$ over the range 5100-6100 \AA\
are listed for each galaxy in Table \ref{continuumtable}.  As
expected, these galaxies have very low polarization, with a median
value of $p = 0.36\%$.  Only one object, NGC 3718, has a continuum
polarization greater than 1\%, and in this case the polarization is
the result of transmission through interstellar dust in the host
galaxy (see \S\ref{section3718}).  Only a small amount of internal or
Galactic dust would be required to produce the low levels of continuum
polarization observed in most of these objects.  Along a given line of
sight, the maximum level of polarization typically produced by
transmission through Galactic dust is $p_{max} (\%) = 9.0$\ebv\
(Serkowski, Mathewson, \& Ford 1975).  Thus, a continuum polarization
of 0.36\% could in principle be produced by transmission through a
reddening column of only \ebv\ = 0.04 mag.

In some galaxies in this sample, $p$ rises steadily toward the blue,
which could be a signature of scattered AGN emission.  If a polarized
nonstellar continuum were present, $p$ would rise toward the blue end
of the spectrum where the nonstellar continuum fraction is greatest.
However, interstellar polarization from Galactic dust or from dust in
the host galaxy can mimic this effect.  As shown by Serkowski \etal
(1975), the wavelength of maximum interstellar polarization ranges
from 4500 to 8000 \AA\ depending on the line of sight.  Since our
wavelength setting does not extend significantly blueward of 4500 \AA,
we are unable to determine whether the $p$ spectrum turns over and
begins to drop off at shorter wavelengths.  Future observations over a
wider wavelength range could result in detections of scattered AGN
emission in some of these objects, if $p$ continues to rise blueward
of 4500 \AA.

Since the magnitude of interstellar polarization is comparable to or
greater than that expected for the intrinsic polarization, special
care must be taken to examine the $q$ and $u$ spectra for polarization
features at \hal.  Faint emission-line features in $q$ or $u$ may not
appear in the rotated Stokes parameter if they are oriented at a
different P.A. from the continuum polarization.  Also, any
emission-line feature at \hal\ appearing in Stokes flux but not in $p$
is most likely a result of interstellar polarization, as it is
generated by multiplying the total flux by a nonzero (but noisy)
quantity. Therefore, only those galaxies having clear emission-line
features in $p$, or in the $q$ or $u$ spectra, are considered to be
detections of physically interesting polarization.  NGC 4395 is the
only exception to this rule, because its nuclear continuum is
predominantly nonstellar; see \S\ref{section4395} below.

To assess the significance level of the emission-line polarizations,
we tested whether the Stokes parameters at \hal\ differ from the
Stokes parameters in the surrounding continuum.  The $q$ parameter was
measured for the total emission (line and continuum) over 6500-6625
\AA, a region which covers the \hal+[\ion{N}{2}] profile.  For the
continuum comparison measurement, a straight line was fitted to the
$q$ spectrum in the line-free regions 6380--6480 \AA\ and 6650--6700
\AA, and the continuum level of $q$ in the \hal\ wavelength bin was
determined from the best-fitting line.  These default line and
continuum wavelength regions were adjusted for a few galaxies with
unusually broad \hal\ emission (NGC 1097, NGC 3998, and M81).  The
quantity $\Delta_q$ gives the significance level, in standard
deviations, of the difference between the value of $q$ measured in the
\hal\ bin and the $q$ continuum level determined from the fit.  The
same procedure was followed for the $u$ spectrum, and Table
\ref{resultstable} lists the results for each galaxy.

Since the $q$ and $u$ spectra commonly show spurious bumps and features
at the level of $1-2\sigma$ with respect to the propagated
uncertainties, we conservatively choose to set a detection threshold
of $4\sigma$ to account for these systematic uncertainties.  Thus, if
$|\Delta_q| \geq 4$ or $|\Delta_u| \geq 4$, we interpret the result as
a significant detection of emission-line polarization. 

The results listed in Table \ref{resultstable} show that three
galaxies in our sample, NGC 315, NGC 1052, and NGC 4261, have
significant detections ($>5\sigma$) of emission-line features in one
Stokes parameter.  In the remaining galaxies, the polarization of the
\hal\ bin is well within 4$\sigma$ of the continuum polarization, and
these results are interpreted as non-detections of emission-line
polarization.  Inspection of the individual polarization spectra,
presented in \S\ref{sectionindividual}, shows that these quantitative
results agree well with visual assessments of detection or
non-detection in the data.

For the type 1 objects, the broad-line flux in total light was
measured by subtracting an absorption-line template spectrum and
deblending the \hal+[\ion{N}{2}] profile, following the methods
described by Ho \etal (1997c).  If broad \hal\ was detected in total
flux but not in $p$, we determined $4\sigma$ upper limits to the
polarization of the broad component under the assumption that the
narrow emission lines are intrinsically unpolarized, but allowing for
the possibility that there may be some intrinsic continuum
polarization.  The quoted limits on $p$(\hal) were calculated by
determining the $4\sigma$ upper limits to the presence of polarization
features in $q$ and $u$ individually.

Throughout this paper, all Galactic reddening estimates refer to the
reddening map of Schlegel, Finkbeiner, \& Davis (1998).

\section{Individual Galaxies}\label{sectionindividual}

\subsection{NGC 315}
\placefigure{ngc315plot}

This elliptical galaxy hosts a broad-lined LINER nucleus (Ho \etal
1997c) surrounded by a dusty circumnuclear disk (\cite{ho98}), and it
is the source of a double-sided radio jet (Giovannini \etal 1990).
Broad features at \hal\ and \hbeta\ are visible in the $q$ and $u$
spectra, but the polarization angles of the continuum and the \hal\
emission are misaligned by $\sim40\arcdeg$.  This offset suggests that
much of the apparent continuum polarization is due to Galactic
foreground dust.  NGC 315 lies at $b = -32\arcdeg$ behind a Galactic
reddening of \ebv\ = 0.065 mag.

To remove the effects of interstellar polarization, we observed the
Galactic A2 star BD+30\arcdeg135 during the January 1998 observing
run.  The star is located at a projected separation of 1\fdg3 from NGC
315, and its apparent magnitude implies a distance of $\approx 530$
pc, large enough for it to be a good probe of Galactic polarization.
The two sight-lines appear to be affected by similar interstellar
polarization: over the range 5100-6100 \AA, NGC 315 has $p = 0.61\%$
at $\theta = 100\arcdeg$, while BD+30\arcdeg135 has $p = 0.59\%$ at
$\theta = 108\arcdeg$.  The Stokes parameter spectra of the star were
fitted with low-order polynomials and the resulting fits were
subtracted from the Stokes parameters of NGC 315, to yield a corrected
set of parameters for the galaxy.  The spectra shown in Figure
\ref{ngc315plot} are the results after subtraction of the interstellar
polarization.

The polarization angle of the \hal+[\ion{N}{2}] blend is $51\arcdeg
\pm 2\arcdeg$, which is offset by $\sim84\arcdeg$ from the radio jet
axis measured at VLBI resolution (\cite{ven93}).  After correction for
interstellar polarization, the continuum over 5100-6100 \AA\ has
$p=0.16\%$ oriented at $\theta = 61\arcdeg$, roughly aligned with the
emission-line polarization angle.

Over the entire \hal+[\ion{N}{2}] blend, the emission-line
polarization is $3.4\% \pm 0.3\%$.  Starlight subtraction of the total
flux spectrum was attempted using several emission-free template
spectra, but none could perfectly reproduce the shape of the stellar
continuum over the 6000-6700 \AA\ range; a spectrum of NGC 3115 was
chosen as the best template.  Ho \etal (1997c) used a seven-Gaussian
decomposition to fit the \hal+[\ion{N}{2}] profile in NGC 315, but due
to the lower spectral resolution of our data and an imperfect
starlight subtraction, we chose to fit the blend using a simpler
quadruple-Gaussian model, representing the three narrow components and
a broad \hal\ line.  The fits in total flux and Stokes flux are shown
in Figure \ref{ngc315hal}.  The broad component of \hal\ has FWHM =
3200 \kms\ in Stokes flux and 2300 \kms\ in total flux.  The ratio of
the two line strengths gives $p \approx 4.5\%$ for broad \hal, but
this measurement is very crude as a result of the noisy \hal\ profile
in polarized light and the uncertain profile decompositions.
\placefigure{ngc315hal}

\subsection{NGC 1052}

Results for NGC 1052 have been presented by Barth \etal (1999a).  The
wings of the broad \hal\ profile are polarized, and the broad \hal\
line has FWHM $\approx2000$ \kms\ in total flux and FWHM $\approx5000$
\kms\ in polarized light.  The angle of polarization is offset by
67\arcdeg\ from the parsec-scale radio axis and by 83\arcdeg\ from the
kpc-scale radio axis, roughly in agreement with expectations for the
obscuring torus model.  We note that this dataset has significantly
higher \sn\ than our observations of the other galaxies in this
sample, owing to the brightness of this galaxy and the relatively long
integration.

\subsection{NGC 1097}
\placefigure{ngc1097plot}

The double-peaked broad \hal\ line which was first discovered by
Storchi-Bergmann, Baldwin, \& Wilson (1993) is still visible in our
January 1998 and January 1999 data (Figure \ref{ngc1097plot}).  The
displayed spectra are weighted averages of the results from these two
observing runs.  Spectropolarimetric observations of such
double-peaked emitters can provide constraints on accretion disk
models for the origin of the broad-line emission (\cite{ch90};
\cite{aha96}; \cite{cht97}).

The \hal\ polarization was evaluated over the wavelength range
6630--6710 \AA, which includes most of the broad double-peaked
feature.  No significant polarization was found, with an upper limit
of $p < 3\%$ for the polarization of the broad \hal\ emission over
this range.  This limit is not stringent enough to constrain accretion
disk models, as the simplest disk-emission model predicts a broad-line
polarization of $p \approx 0.5\%$ (\cite{cha60}; \cite{ch90}) for a
disk inclination of 34\arcdeg, the orientation which best matches the
double-peaked profile (\cite{sb97}).

\subsection{NGC 1167}
\placefigure{ngc1167plot}

Due to cloudy conditions and the faintness of this AGN, our
observations of NGC 1167 (Figure \ref{ngc1167plot}) have considerably
poorer \sn\ than the other galaxies in the sample.  No polarization
features are seen at \hal.

\subsection{NGC 1667}
\placefigure{ngc1667plot}

Although there is a faint bump in the $p$ spectrum near the wavelength
of \hal\ (Figure \ref{ngc1667plot}), inspection of the Stokes
parameter spectra shows that this feature consists of a narrow bump on
the red side of the \hal+[\ion{N}{2}] profile.  The Stokes parameters
in the \hal\ wavelength bin do not deviate significantly from their
values in the surrounding continuum.  We conclude that polarized \hal\
emission has not been detected, but deeper observations would be
worthwhile.

\subsection{NGC 1961}
\placefigure{ngc1961plot}

NGC 1961 is located at a Galactic latitude of 19\arcdeg, and its
continuum appears heavily reddened (Figure \ref{ngc1961plot}).  Its
nucleus has a low polarization of $0.03\% \pm 0.02\%$ over 5100-6100
\AA, despite the Galactic reddening of \ebv\ = 0.12 mag along this
line of sight.

\subsection{NGC 2639}
\placefigure{ngc2639plot}

NGC 2639 contains H$_2$O megamasers similar to those seen in NGC 4258
(\cite{wbh95}), suggesting that an accretion disk is present.  The
narrow emission lines in NGC 4258 are polarized at the level of $p =
2-12\%$ (\cite{wil95}; \cite{bar99b}), but no such features are found
in NGC 2639 (Figure \ref{ngc2639plot}).  The broad component of \hal\
is polarized at a level of $<4\%$.  The continuum polarization rises
to the blue, and may be the result of dust transmission within NGC
2639, as the Galactic reddening of \ebv\ = 0.025 mag is by itself
probably insufficient to produce a polarization of $\sim0.5\%$.
Further observations shortward of 4500 \AA\ could determine whether
this continuum polarization represents a featureless AGN emission
component or is the result of transmission of the starlight spectrum
through foreground dust.

\subsection{M81 (NGC 3031)}
\placefigure{m81plot}

Antonucci (1993) suggested that the broad \hal\ emission in M81 might
be seen in reflection, as the apparent lack of variability of the
broad-line emission prior to 1993 (\cite{hfsm81}) could be interpreted
as the result of the light-travel time across a spatially extended
reflection region.  No broad-line polarization is detected at \hal,
however, with an upper limit of $p < 1.3\%$ for the broad component of
\hal\ over the range 6500--6700 \AA.  The broad double-peaked \hal\
emission first detected in 1995 (\cite{bow96}) appears to have become
by January 1999 primarily an excess in the red wing (see Figure
\ref{m81plot}), while the blue wing, which extended to 6400 \AA\ in
the 1995 spectra, has apparently faded.

\subsection{NGC 3642}
\placefigure{ngc3642plot}

This LINER has a broad \hal\ emission component (\cite{hfsp97}) and a
spatially unresolved nuclear ultraviolet source (\cite{bar98}).  No
polarization features are seen at \hal\ (Figure \ref{ngc3642plot}).
The broad \hal\ component is faint in total flux and we are only able
to set a relatively high limit to its polarization of $p < 4.6\%$.

\subsection{NGC 3718}\label{section3718}
\placefigure{ngc3718plot}

The nuclear continuum of NGC 3718 has $p \approx 4-6\%$.  This
polarization is presumably due to foreground dust within NGC 3718, as
the galaxy lies behind a small Galactic dust column with reddening of
only \ebv\ = 0.014 mag.  Comparison of the continuum shape with
spectra of unreddened galaxies suggests a total reddening of roughly
\ebv\ $\approx 1$ mag.  An \hst\ $V$-band image of NGC 3718 shows that
its nucleus is extremely dusty (Barth \etal 1998).

In both $q$ and $u$, the polarization in the \hal\ bin differs from
that of the surrounding continuum by $\sim 2.8\sigma$, but inspection
of the Stokes parameter spectra shows that there are variations at
comparable significance levels at other wavelengths as well.  Under
the assumption that the continuum polarization is entirely due to dust
transmission, the broad component of \hal\ has $p <3\%$.

\subsection{NGC 3998}
\placefigure{ngc3998plot}

NGC 3998 has a bright, low-ionization, narrow-line spectrum with
prominent broad wings on the \hal\ line (Figure \ref{ngc3998plot}).
No features are detected in $q$ or $u$, and we set an upper limit of
1.4\% to the polarization of the broad \hal\ component.

\subsection{NGC 4261}
\placefigure{ngc4261plot}

This elliptical galaxy hosts an impressive double-lobed radio jet
oriented approximately east-west (PA = $88\arcdeg \pm 1\arcdeg$;
Birkinshaw \& Davies 1985) and a dusty circumnuclear disk with a
diameter of 120 pc (\cite{jaf93}).  Ferrarese \etal (1996) analyzed
the velocity field of the disk using \hst\ spectra and found Keplerian
rotation about a central dark mass inferred to be $(4.9 \pm 1.0)
\times 10^8$ \msun.

The \hal\ emission line is very weakly polarized, as seen in Figure
\ref{ngc4261plot}; the entire \hal+[\ion{N}{2}] blend has $p = 1.4\%
\pm 0.3\%$ over 6500--6625 \AA\ (rest wavelength).  The emission-line
polarization is aligned at PA = $6.5\arcdeg \pm 5\arcdeg$, which is
offset by 81.5\arcdeg\ from the radio axis.  The major axis of the
disk is at $-16\arcdeg$ (Jaffe \etal 1993), so the polarization is
offset by 22.5\arcdeg\ from the disk plane.

In total flux, the \hal+[\ion{N}{2}] profile appears to have broad
wings, but careful inspection of the spectrum reveals that similarly
broad wings may be present in the [\ion{S}{2}] lines as well.  It is
possible to fit the \hal+[\ion{N}{2}] profile reasonably well either
with or without a broad \hal\ component, and we are unable to reject
the null hypothesis that no broad-line component is present. This
conclusion is in agreement with the results of Jaffe \etal (1996) and
Ho \etal (1997c).  In polarized flux, the spectrum is too noisy to
allow a meaningful decomposition into broad and narrow components.  A
single-Gaussian fit to the \hal+[\ion{N}{2}] profile gives FWHM = 2800
\kms, but deeper observations would be needed to confirm the width of
this feature.

\subsection{NGC 4395}\label{section4395}
\placefigure{ngc4395plot}

Ground-based and \hst\ images and spectra of this low-luminosity
Seyfert 1 galaxy have revealed the presence of a compact source of
optical continuum emission, with a FWHM size of $\lesssim0.7$ pc, as
well as a featureless ultraviolet continuum and high-excitation broad
and narrow emission lines (\cite{fs89}; \cite{fhs93}).  NGC 4395 is
unique among the galaxies observed in this sample in that it is the
only one in which the nonstellar continuum is clearly visible in the
optical spectrum, and no stellar absorption features are detected.  In
this case, if the continuum and broad emission lines were viewed
through the same scattering geometry, no broad-line feature in $p$
would be expected.  Instead, the signature of a genuinely polarized
AGN continuum would be a drop in $p$ at the wavelengths of the narrow
emission lines, assuming these lines to be intrinsically unpolarized.
This effect is seen, for example, in the Seyfert 1 galaxy Mrk 509
(\cite{gm94}).

The continuum polarization is $0.67\% \pm 0.03\%$, with $\theta =
30\arcdeg \pm 2 \arcdeg$, over 5100--6100 \AA.  This polarization is
unlikely to be the result of Galactic foreground dust, as the Galactic
reddening to NGC 4395 is only \ebv = 0.017 mag.  Furthermore, Galactic
interstellar polarization would affect all of the emission lines and
and the continuum equally, but Figure \ref{ngc4395plot} shows that
this is not the case.  \hal\ (and probably \hbeta) appear in Stokes
flux while the [\ion{O}{3}] $\lambda\lambda4959, 5007$ lines do not,
other than the residuals that result from multiplying the extremely
narrow line profiles in total flux by the random noise in the $p$
spectrum.  Thus, the polarization must be occurring within NGC 4395.

Figure \ref{ngc4395hal} shows a comparison of the \hal+[\ion{N}{2}]
profiles in total flux and Stokes flux.  Although the profile is noisy
in polarized light, it appears that the broad and narrow components of
\hal\ are present in Stokes flux, while [\ion{N}{2}] \lam6584 is not.
Transmission through foreground dust within the nucleus of NGC 4395
could be responsible for the polarization, but only if the region of
aligned dust grains covers the continuum and BLR but not the more
spatially extended NLR.  Electron scattering of the continuum and
broad lines is a more likely explanation, and consistent with the
relatively flat wavelength dependence of $p$ in the continuum.
Comparison of $\theta$ with a radio axis would provide clues to the
scattering geometry, but the nuclear radio source in NGC 4395 is
unresolved (\cite{sra92}).  A clear sign of electron scattering would
be a broadened \hal\ profile in Stokes flux, but data of higher \sn\
and greater spectral resolution would be required in order to deblend
the \hal+[\ion{N}{2}] emission convincingly, because the broad \hal\
line has a very non-Gaussian profile (\cite{hfsp97}).

\subsection{NGC 4594}\label{section4594}
\placefigure{ngc4594plot}

Kormendy \etal (1996) detected a faint broad component of \hal\ with
FWZI $\approx 5200$ \kms\ in \emph{HST} spectra.  No broad-line
polarization is detected, but the continuum polarization rises
gradually toward the blue end of the spectrum (Figure
\ref{ngc4594plot}), from 0.40\% over 6500--6800 \AA\ to 0.57\% over
4500--5000 \AA.  Much of this continuum polarization could be the
result of transmission through Galactic dust, since the reddening
column along this line of sight is \ebv\ = 0.05 mag.

\section{Discussion}

In understanding the place of LINERs within AGN unified schemes, a
question of paramount importance is whether the broad \hal\ wings
observed in some LINERs are viewed directly, or in light scattered
above the midplane of an obscuring torus.  In other words, as
Antonucci (1993) has asked, are these LINER 1.9 nuclei intrinsically
type 1 or type 2 objects?  One argument, based on the obscuring torus
model, can be given as follows.  Suppose that the LINER 1.9 nuclei are
in fact type 2 objects, in which the broad \hal\ emission is seen only
in scattered light.  If this were the case, then there should be lines
of sight from which these galaxies would appear as type 1 objects, and
the broad-line emission would appear brighter by perhaps two orders of
magnitude or more.  Thus, unless the torus opening angle is very small
in these objects, a significant fraction of LINER nuclei should have
broad-line equivalent widths of at least 100 \AA, as compared with the
typical equivalent widths of a few \AA\ observed in LINER 1.9 nuclei
(\cite{hfsp97}).  But very few such nuclei are known to exist; in the
Ho \etal survey of 486 nearby galaxies, not a single LINER was
found to have a broad-line classification in the 1.0--1.8 range, under
the Osterbrock (1977; 1981) classification scheme.  The few known
examples of LINERs with very high equivalent width broad \hal\ lines
include NGC 7213 (\cite{fh84}), the radio galaxy Pictor A
(\cite{fil85}), and the transient double-peaked source in NGC 1097
(\cite{sbw93}).

The fact that LINER 1.0--1.8 nuclei are so rare provides a strong
suggestion that the LINER 1.9 nuclei \emph{are} those LINERs in which
we have a direct view of the nucleus, and that the broad emission
lines in these objects are intrinsically faint.  The undetectably low
emission-line polarizations observed in most LINER 1.9 nuclei are
consistent with this hypothesis.  If this line of reasoning is
correct, then the LINER 1.9 nuclei NGC 315 and NGC 1052, in which
polarized broad \hal\ is detected, may be cases in which the BLR is
viewed both directly and in reflected light.  This explanation can
also account for the low polarizations of the broad \hal\ lines in
these galaxies, in comparison with the much higher broad-line
polarizations of 15--35\% measured in some Seyfert 2 nuclei
(\cite{tra95}).

Scattering within the opening cone of (or above the plane of) an
obscuring torus provides a consistent explanation for the
emission-line polarization observed in NGC 315, NGC 1052, and NGC
4261, since the angle of polarization in each case is roughly
perpendicular to the axis of the radio jet.  The nuclear dust disks
seen in \emph{HST} images of NGC 4261 (\cite{jaf93}) and NGC 315
(\cite{ho98}) provide supporting evidence for this interpretation, as
these disks are likely to be the outer extensions of the obscuring
tori.  The nuclear morphology of NGC 1052 has yet to be studied with
\emph{HST} data, but we note that ground-based observations have
detected rotating disks of ionized (\cite{di86}) and neutral
(\cite{vg86}) gas on kiloparsec scales which could be connected with a
dusty torus and/or accretion disk on smaller scales.

The scattering material may consist of either dust particles or
electrons, and it is difficult to distinguish between these
possibilities with the limited information contained in our data.  In
NGC 1052, the broad \hal\ emission is considerably broader in
polarized light than in total light (5000 \kms\ vs. 2000 \kms), and
this difference can be interpreted as the result of electron
scattering by a $\sim10^5$ K scattering medium, if the BLR is viewed
partially in direct light as well as in scattered light.  In general,
the best way to discriminate between dust and electron scattering is
to obtain spectropolarimetric observations in the ultraviolet, where a
$p$ spectrum rising to the blue would be a clear indicator of dust
scattering rather than wavelength-independent Thomson scattering.
Ultraviolet spectropolarimetry would also alleviate the problem of
contamination by the surrounding old stellar population.

One curious result of this survey is that polarized emission lines
have been detected in all three of the ellipticals that were observed,
but in none of the spirals.  While it is true that our sample was not
selected in a statistically meaningful way, this dichotomy between
ellipticals and spirals is certainly intriguing, and should be
investigated further with a larger, carefully selected sample of
LLAGNs.  It will probably be very difficult to detect polarized
broad-line emission from LINER 2 nuclei in spiral hosts, however,
since these nuclei tend to be very dusty (\cite{bar98}) and the
emission-line equivalent widths tend to be low.  It may be easier to
detect hidden BLRs in elliptical LINERs simply because these objects
are less likely to be obscured by foreground dust lanes.

Heisler \etal (1997) have shown that Seyfert 2 nuclei with hidden BLRs
have systematically warmer far-infrared colors, as measured by the
\iras\ $25 \micron / 60 \micron$ flux ratio, than do Seyfert 2
galaxies without hidden BLRs.  They interpret this trend as the result
of viewing angle: in their model, galaxies in which hidden BLRs are
detected are those where our line of sight grazes the inner edge of
the torus, and we view directly the hot dust in the inner torus.
Determining whether the infrared colors of LINERs follow this trend as
well would require infrared data of higher spatial resolution, since
the \iras\ beamsize includes much of the host galaxy as well as the
faint nuclear emission.  It is interesting that the two galaxies in
our sample with the warmest \iras\ colors are NGC 1052 and NGC 4261,
with $f_{25}/f_{60}$ = 0.57 and 1.00, respectively, although the
measurement for NGC 4261 carries an 86\% uncertainty (using the
results tabulated by \cite{hfs97params}).  For the full sample of 14
galaxies observed in this survey, the mean $f_{25}/f_{60}$ color is
0.27.  NGC 315 was not detected at 25 \micron, resulting in an upper
limit of $f_{25}/f_{60} < 0.47$.

The question of whether \emph{all} Seyfert 2 galaxies contain hidden
type 1 nuclei has been difficult to address partly because of the
starlight contamination problem.  Among Seyfert 2 nuclei, there is a
strong tendency for hidden BLRs to be detected preferentially in
objects where the contamination by unpolarized starlight is low
(\cite{km98}).  Most Seyfert 2 galaxies known to have hidden BLRs have
galaxy fractions of $F_g < 0.5$ (\cite{tra95}), where $F_g$ is the
fraction of the optical continuum flux contributed by starlight.  Our
results indicate that some LINERs, in which $F_g > 0.9$, contain
polarized broad-line emission which is detected at the level of only a
few tenths of a percent in $p$ above the level of the surrounding
continuum.  Many Seyfert 2 nuclei with hidden BLRs have probably been
missed in previous surveys, which may not have been able to detect
such faint polarization features.  To alleviate the starlight problem,
it would be useful to conduct a spectropolarimetric survey of Seyfert
2 nuclei at \sn\ comparable to those we have reached in this work,
with a sample which is not biased toward objects of high polarization.
As shown by Kay \& Moran (1998), detection of hidden BLRs in
starlight-dominated Seyfert 2 nuclei could provide important
constraints on the applicability of the obscuring torus model to the
Seyfert population as a whole.

\section{Conclusions}

Our primary conclusion is that polarized broad-line emission is seen
in some nearby galaxies classified as LINERs, indicating that unified
models of AGNs apply to at least some members of this class.  Most
galaxies we observed do not show emission-line polarization, however.
A likely explanation is that most LINERs in which broad \hal\ wings
are detected in total flux are intrinsically ``type 1'' objects in
which we view the central regions of the AGN directly.  Objects
classified as ``type 2'' LINERs may have hidden BLRs, as in NGC 4261,
and further observations should be able to detect more objects of this
type and to constrain the fraction of LINER 2s having hidden BLRs.

The objects in which we have detected polarized broad \hal\ are all
elliptical galaxies with double-sided radio jets, and these may be
interpreted as low-luminosity versions of more powerful narrow-line
radio galaxies, some of which are known to host hidden quasars
(\cite{ab90}).  Polarized broad-line emission was not detected in any
LINERs in spiral hosts.  As LINER 2 nuclei may be the most common
manifestation of the AGN phenomenon, further searches for hidden BLRs
in these objects can play an important role in determining the
fraction of galaxies in which accretion-powered nuclear activity
occurs.

\acknowledgements

We are grateful to Luis Ho for providing the template galaxy spectra
used in this work.  An anonymous referee provided helpful suggestions
which improved the analysis and presentation of our data.  The
W. M. Keck Observatory is operated as a scientific partnership among
the California Institute of Technology, the University of California,
and NASA, and was made possible by the generous financial support of
the W.M. Keck Foundation.  This work was supported by NASA grant NAG
5-3556.  Research by A. J. B. is supported by a postdoctoral
fellowship from the Harvard-Smithsonian Center for Astrophysics.
E. C. M.  acknowledges partial support by NASA through Chandra
Fellowship grant PF8-10004 awarded by the Chandra X-ray Center, which
is operated by the Smithsonian Astrophysical Observatory for NASA
under contract NAS8-39073.

\clearpage

\begin{deluxetable}{lcccccl}
\tablecaption{Journal of Observations\label{journal}} 
\tablehead{\colhead{Galaxy} & \colhead{Type} & \colhead{Nucleus} &
\colhead{log [$L$(\hal)/(erg s\per)]} & \colhead{Exposure (s)} &
\colhead{UT Date} & \colhead{notes}}
\startdata
NGC 315  & E   & L1.9  & 39.6(a) &  3600 & 20 Dec 97 &  \nl
NGC 1052 & E   & L1.9  & 40.1(b)  & 2700 & 20 Dec 97 &  \nl
NGC 1097 & SBb & L1.9  & 39.2(c) &  2000 & 17 Jan 98 & thin cirrus \nl
         &     &       &       &  2400 & 06 Jan 99 &  \nl
NGC 1167 & S0  & S2    & 40.2(a) &  2400 & 17 Jan 98 & thin cirrus  \nl
NGC 1667 & Sc  & S2    & 40.9(d) & 3600 & 07 Mar 98 &  \nl
NGC 1961 & Sc  & L2    & 40.4(e) & 2800 & 20 Dec 97 &  \nl
NGC 2639 & Sa  & S1.9  & 40.5(c) & 3600 & 06 Jan 99 & cirrus \nl
NGC 3031 & Sab & S1.5  & 38.6(c) & 600  & 06 Jan 99 & cirrus  \nl
NGC 3642 & Sbc & L1.9  & 39.9(c) & 3600 & 07 Mar 98 &  \nl
NGC 3718 & SBa & L1.9  & 38.5(a) & 3600 & 17 Jan 98 & thick clouds \nl
NGC 3998 & S0  & L1.9  & 40.0(a)  & 600  & 06 Jan 99 & thick cirrus \nl
NGC 4261 & E   & L2    & 39.4(a) &  2400 & 20 Dec 97 &  \nl
NGC 4395 & Sm  & S1.8  & 38.1(f) & 2400 & 17 Jan 98 & clouds \nl
NGC 4594 & Sa  & L2    & 39.7(a) &  1000 & 06 Jan 99 & cirrus  \nl
\enddata

\tablecomments{Morphological types are taken from the NASA
Extragalactic Database (NED).  Spectroscopic types are from Ho \etal
(1997a); S = Seyfert and L = LINER.  The numerical classification
indicates the relative detectability of the broad \hal\ and \hbeta\
lines, according to the system of Osterbrock (1977; 1981).  Narrow
\hal\ luminosities are from (a) Ho \etal (1997a); (b) Heckman (1980);
(c) Keel (1983); (d) Radovich \& Rafanelli (1996); (e) this work; (f)
Kraemer \etal (1999).  For consistency, the luminosities are scaled to
the distances given by Ho \etal (1997a).}
\end{deluxetable}

\clearpage

\begin{deluxetable}{lcc}
\tablecaption{Continuum Polarization Results\label{continuumtable}}

\tablehead{ \colhead{Galaxy} & \colhead{$p$(5100-6100 \AA) (\%)} &
\colhead{$\theta$ (\arcdeg)} }
\startdata

NGC 315\tablenotemark{a} & $0.16 \pm 0.01$ & $60 \pm 2$ \nl
NGC 1052 & $0.43 \pm 0.01$ & $\phn5 \pm 1$ \nl
NGC 1097 & $0.26 \pm 0.02$ & $178 \pm 2\phn$ \nl
NGC 1167 & $0.80 \pm 0.04$ & $164 \pm 2\phn$ \nl
NGC 1667 & $0.35 \pm 0.02$ & $94 \pm 1$ \nl
NGC 1961 & $0.03 \pm 0.02$ & $152 \pm 25$ \nl
NGC 2639 & $0.56 \pm 0.01$ & $46 \pm 1$ \nl
NGC 3031 & $0.36 \pm 0.01$ & $35 \pm 1$ \nl
NGC 3642 & $0.20 \pm 0.01$ & $17 \pm 2$ \nl
NGC 3718 & $4.89 \pm 0.03$ & $138 \pm 1\phn$ \nl
NGC 3998 & $0.11 \pm 0.02$ & $61 \pm 4$ \nl
NGC 4261 & $0.02 \pm 0.01$ & $\phn20 \pm 13$ \nl
NGC 4395 & $0.67 \pm 0.03$ & $30 \pm 2$ \nl
NGC 4594 & $0.49 \pm 0.02$ & $29 \pm 1$ \nl
\enddata
\tablenotetext{a}{After correction for Galactic interstellar polarization.}
\end{deluxetable}

\clearpage

\begin{deluxetable}{lrr}
\tablecaption{\hal\ Polarization Results\label{resultstable}}
\tablehead{\colhead{Galaxy} &  
\colhead{$\Delta_q$} & \colhead{$\Delta_u$} }
\startdata
NGC 315\tablenotemark{a} & $-$2.7 & 11.8 \nl
NGC 1052 & 8.3 & $-$0.8 \nl
NGC 1097\tablenotemark{b} & 3.3 & 1.5 \nl
NGC 1167 & 3.2 & 1.6 \nl
NGC 1667 & 1.1 & $-$1.2 \nl
NGC 1961 & 0.3 & 0.9 \nl
NGC 2639 & 0.0 & $-$1.3 \nl
NGC 3031\tablenotemark{c} & $-$1.2 & $-$0.4 \nl
NGC 3642 & 1.8 & $-$1.3 \nl
NGC 3718 & $-$2.9 & $-$2.8 \nl
NGC 3998\tablenotemark{d} & 1.7 & 2.0 \nl
NGC 4261 & 5.1 & 1.3 \nl
NGC 4395 & 0.8 & $-$0.8 \nl
NGC 4594 & 0.2 & $-$1.9 
\enddata

\tablecomments{The quantities $\Delta_q$ and $\Delta_u$ are measures
of the significance levels for the detection of features at \hal\ in
the $q$ and $u$ spectra, as described in the text. For example, NGC
315 has an $11.8\sigma$ detection of a feature at \hal\ in $u$, and
NGC 1052 has an $8.3\sigma$ detection in $q$.  The regions used for
the continuum fit are 6380--6480 \AA\ and 6650--6700 \AA, and the
\hal\ region is 6500--6625 \AA, except as noted below.  All
measurements refer to rest wavelengths.}

\tablenotetext{a}{After correction for Galactic interstellar
polarization.}
\tablenotetext{b}{For NGC 1097, the \hal\ bin is 6350--6710 \AA\ and
the continuum regions are 6150--6250 \AA\ and 6750--6830 \AA.}
\tablenotetext{c}{For NGC 3031, the \hal\ bin is 6500--6700 \AA\ and
the continuum regions are 6380--6480 \AA\ and 6750--6830 \AA.}
\tablenotetext{d}{For NGC 3998, the \hal\ bin is 6500--6650 \AA\ and
the continuum regions are 6380--6480 \AA\ and 6750--6830 \AA.}

\end{deluxetable}

\clearpage

\begin{center}
\textbf{Figure Captions}
\end{center}

\figcaption[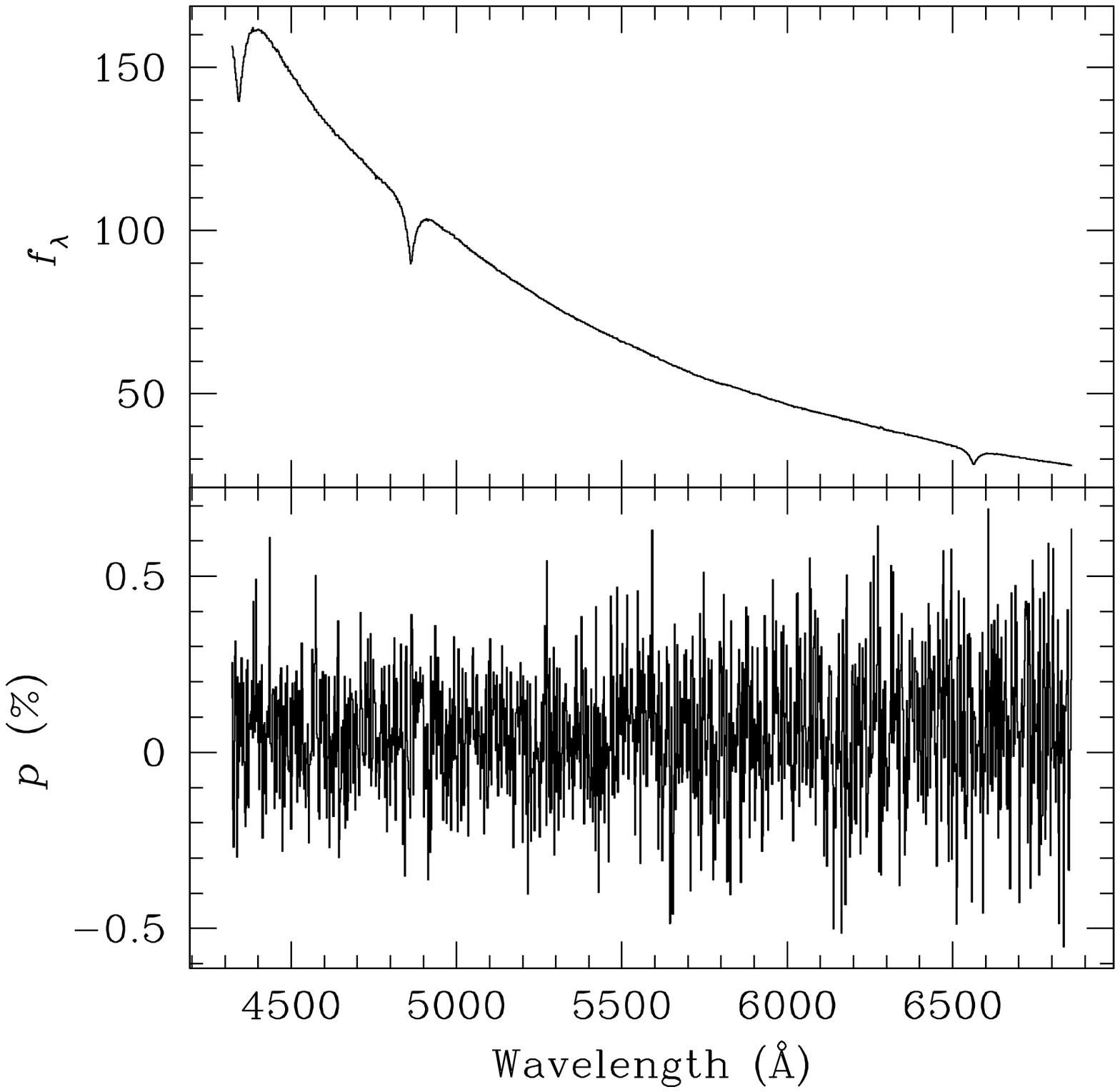]{Total flux spectrum, in units of $10^{-15}$ erg
cm\persq\ s\per\ \AA\per, and polarization of the null standard star
G191B2B.  The average polarization of $0.05\% \pm 0.01\%$ over the
range 4320-5000 \AA\ is in agreement with the $B$-band polarization of
$p = 0.090\% \pm 0.048\%$ measured by Turnshek \etal (1990).
\label{null}}

\figcaption[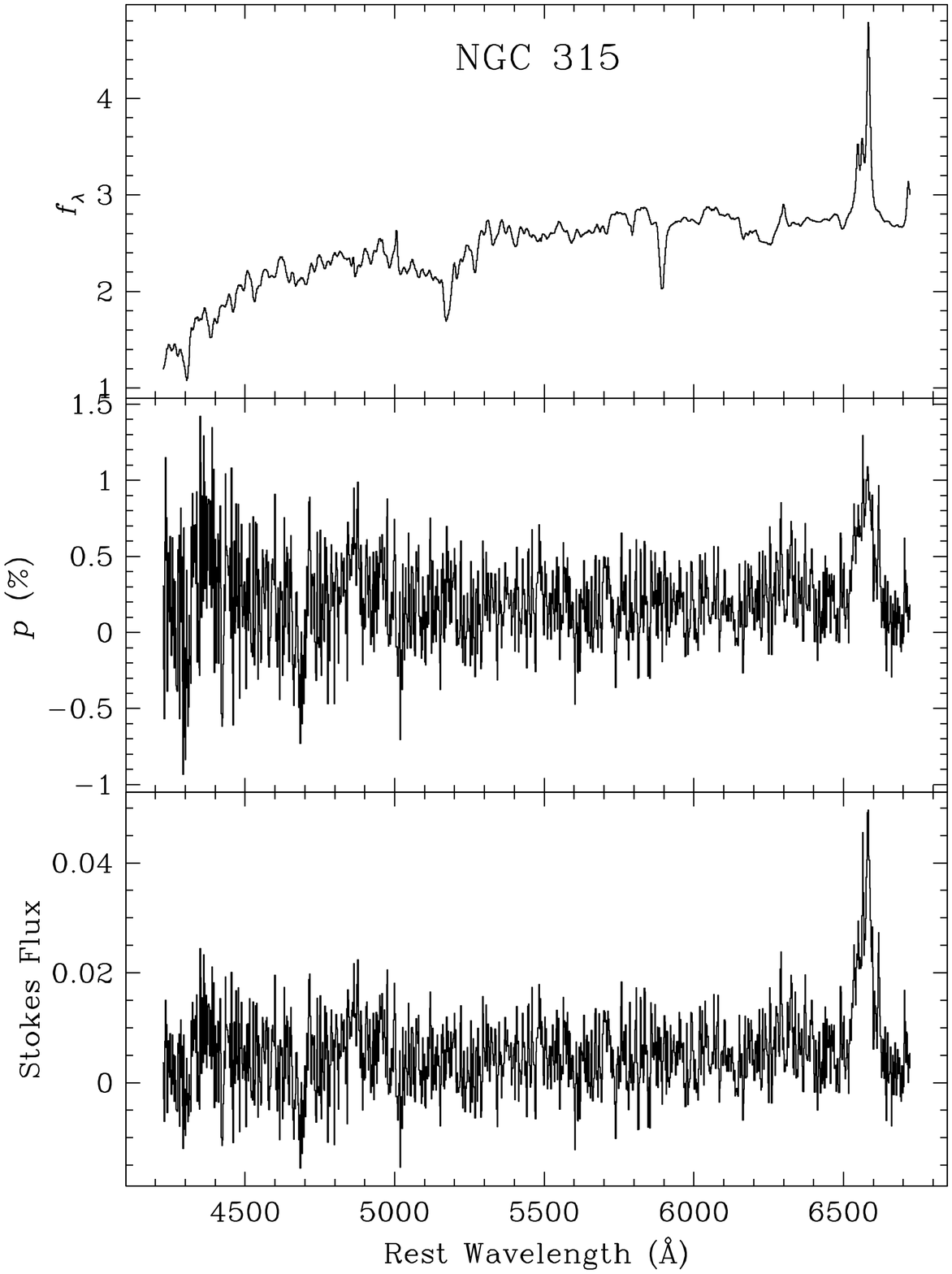]{NGC 315. The data have been corrected for
interstellar polarization by subtracting the Stokes parameters of the
star BD+30\arcdeg135.  Here and in subsequent figures of this type,
the panels displayed are: \emph{Top panel---} Total flux, in units of
$10^{-15}$ erg cm\persq\ s\per\ \AA\per.  \emph{Middle panel---}
Degree of linear polarization, given as the rotated Stokes parameter.
\emph{Bottom panel---} Stokes flux, the product of total flux and the
rotated Stokes parameter.
\label{ngc315plot}}

\figcaption[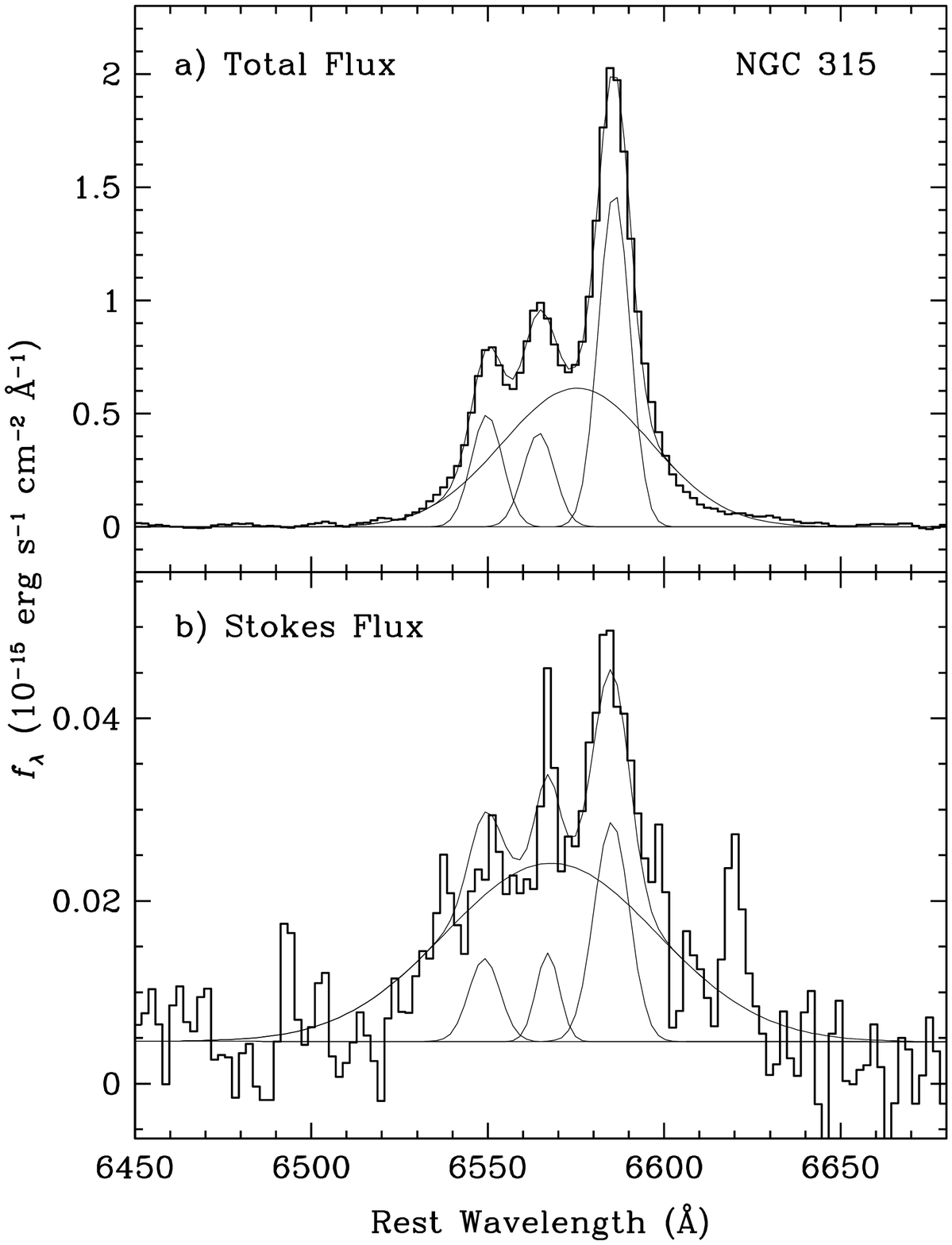]{Comparison of the \hal+[\ion{N}{2}] profiles of NGC
315 in total flux and Stokes flux.  \emph{(a)} Total flux spectrum,
after starlight subtraction.  \emph{(b)} Stokes flux, equal to $p
\times f_\lambda$.
\label{ngc315hal}}

\figcaption[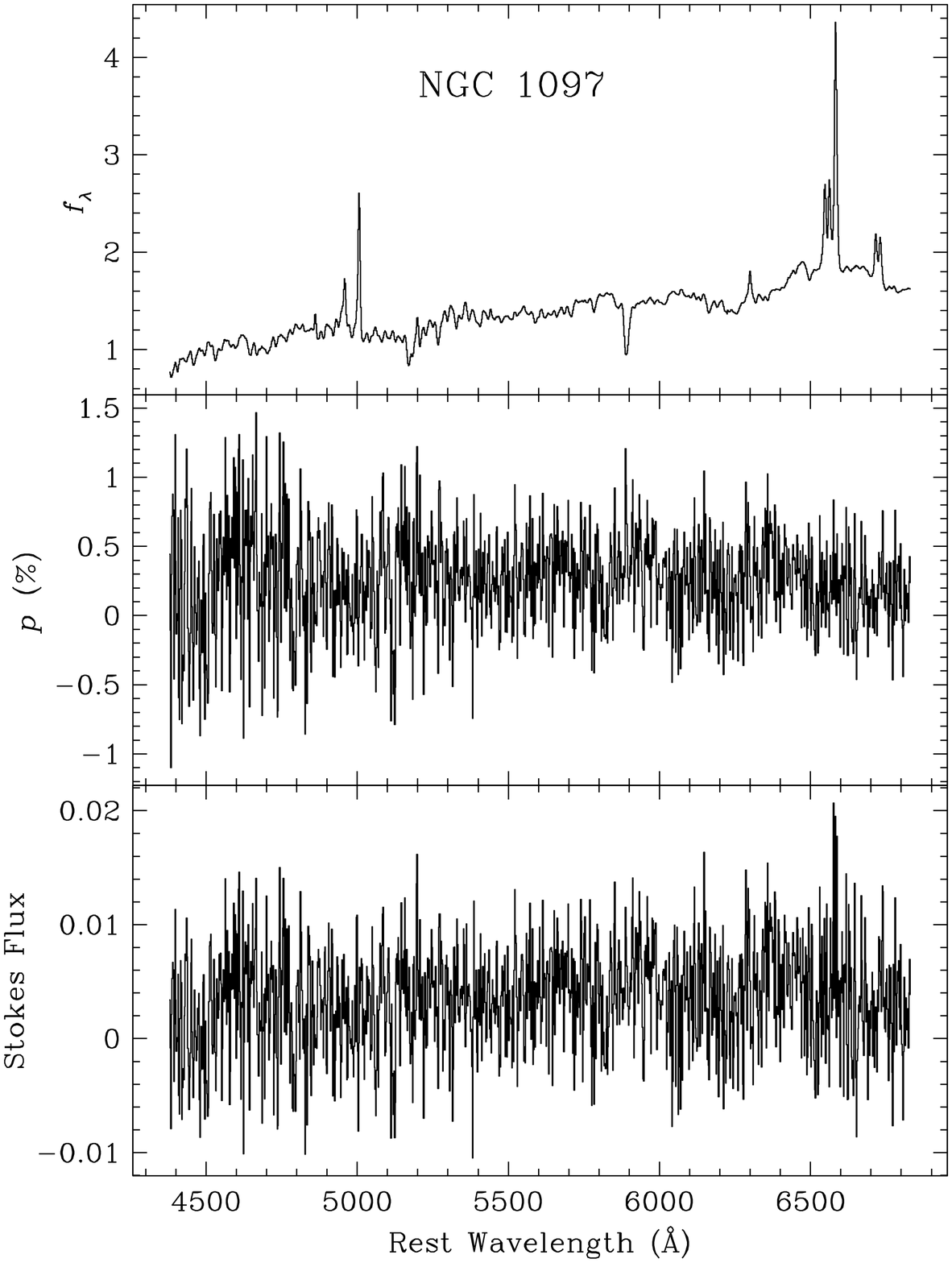]{NGC 1097.  The displayed spectra are weighted
averages of the January 1998 and January 1999
datasets. \label{ngc1097plot}}

\figcaption[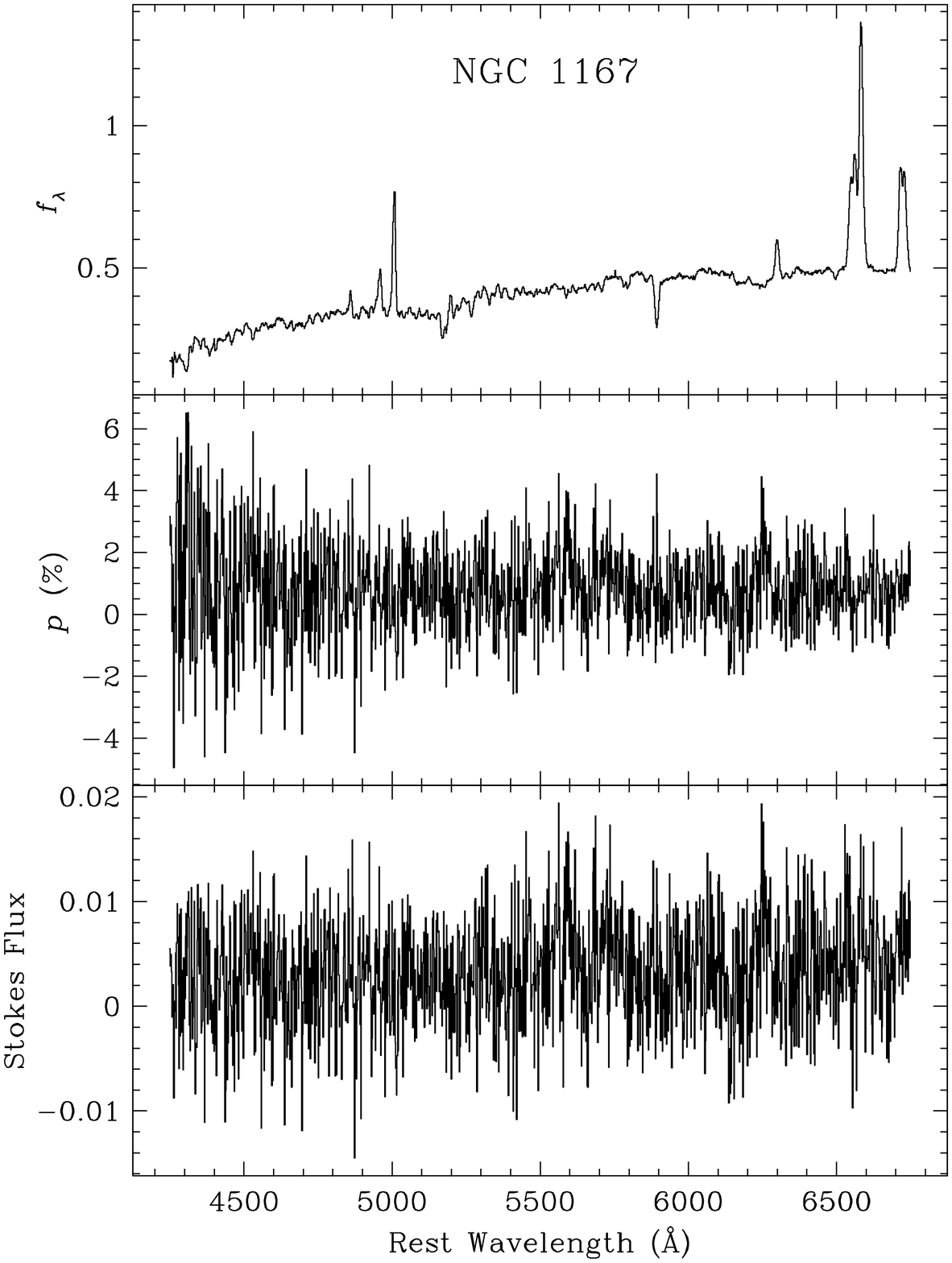]{NGC 1167. \label{ngc1167plot}}

\figcaption[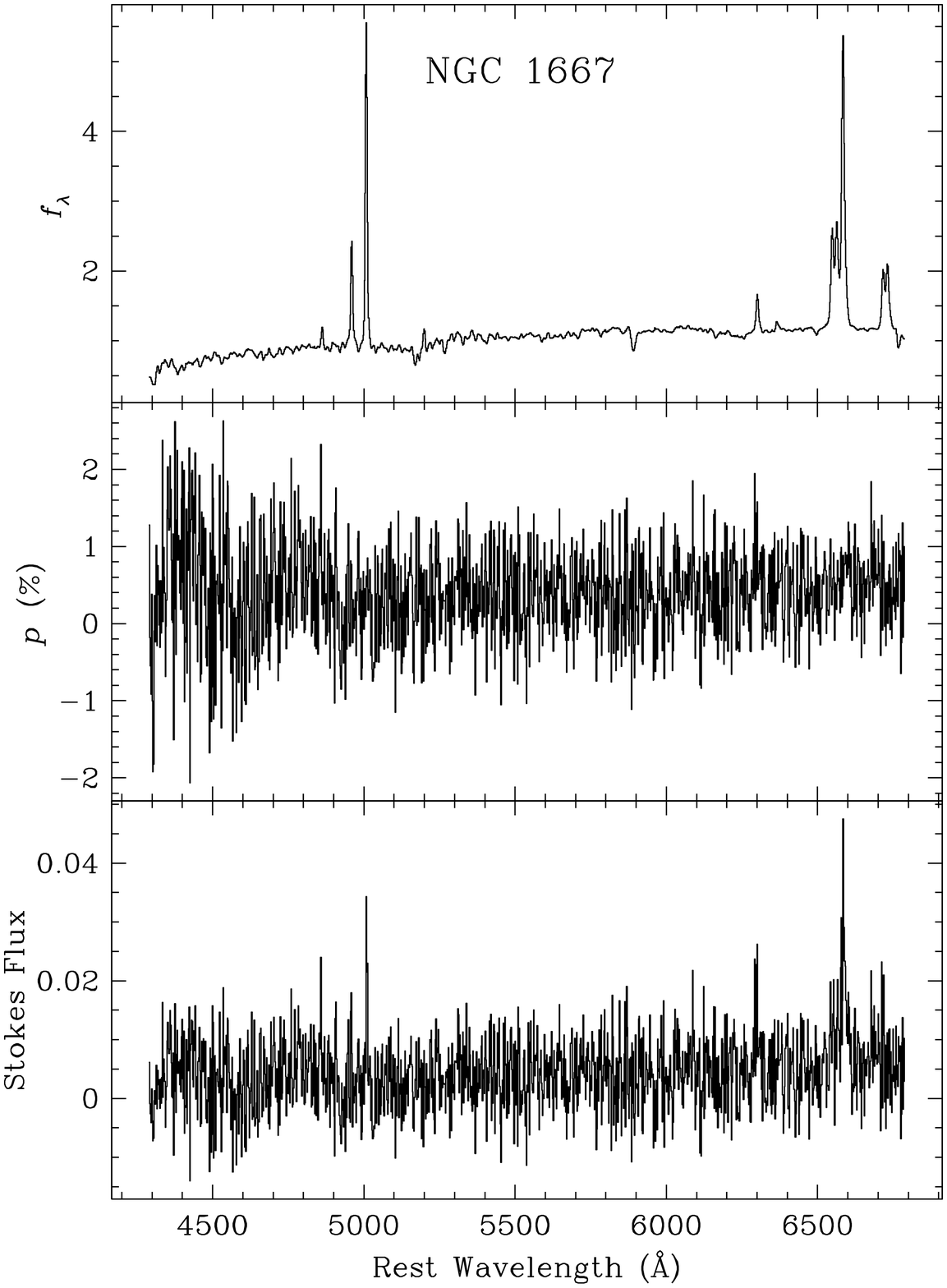]{NGC 1667. \label{ngc1667plot}}

\figcaption[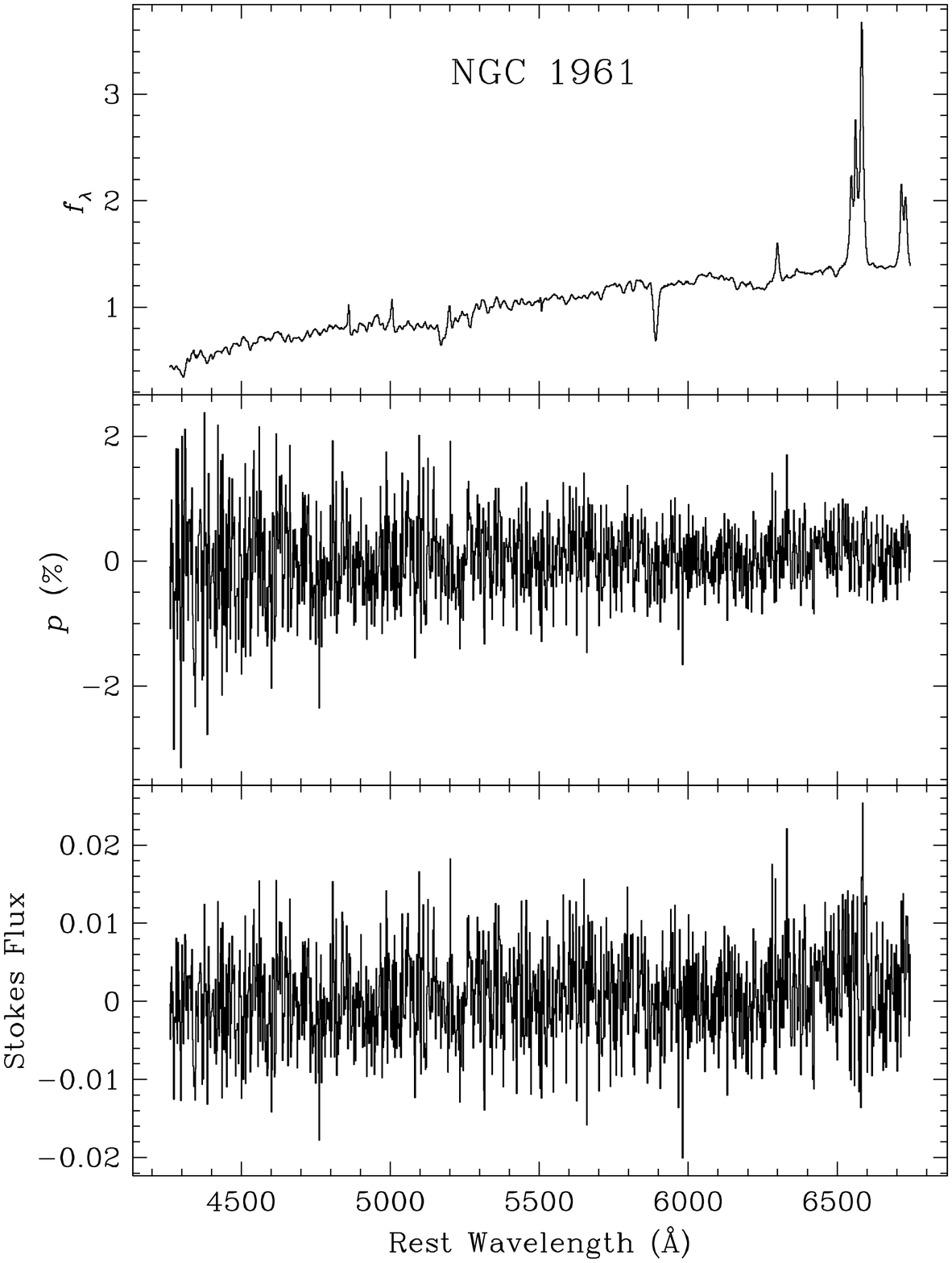]{NGC 1961. \label{ngc1961plot}}

\figcaption[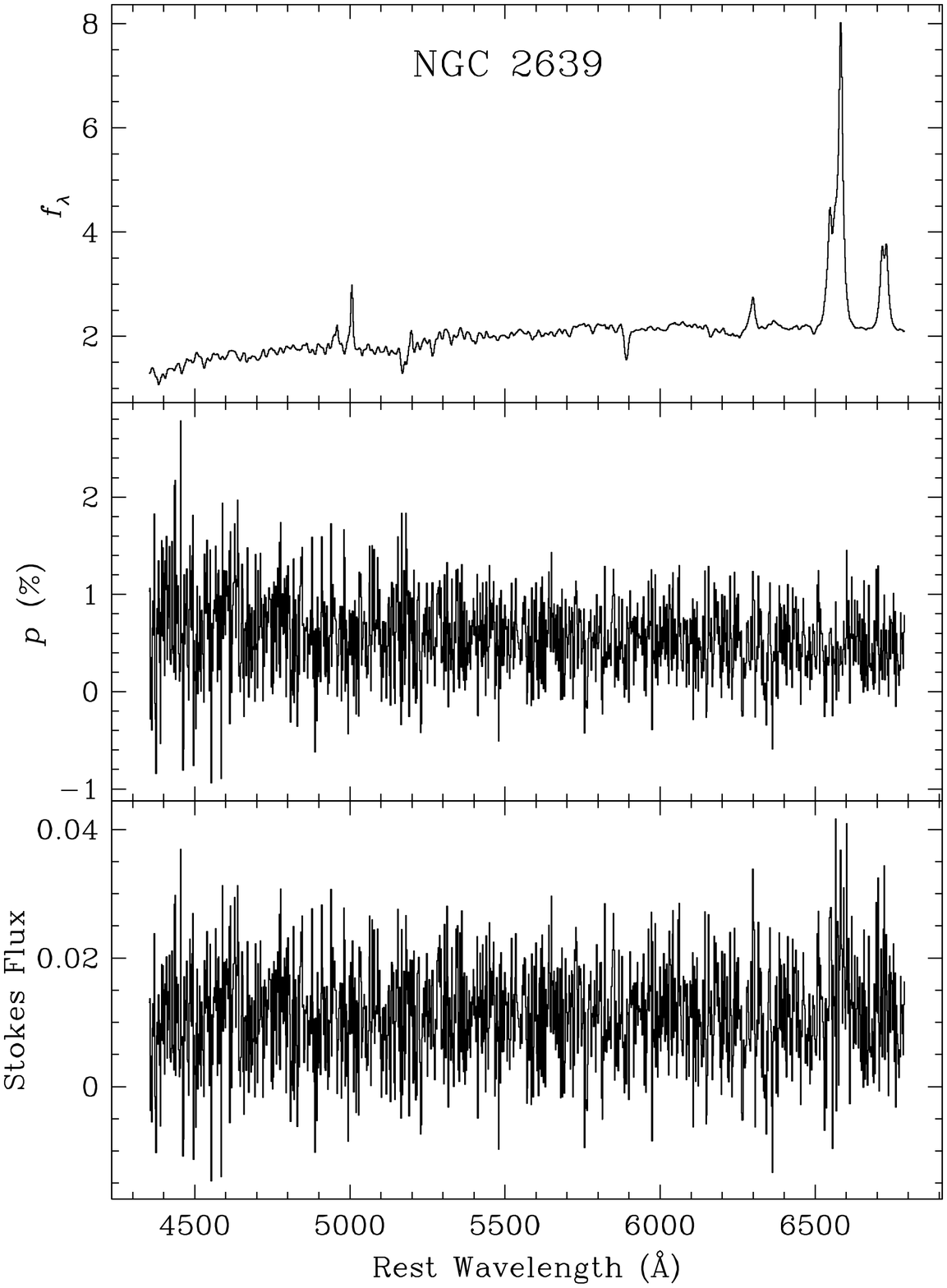]{NGC 2639. \label{ngc2639plot}}

\figcaption[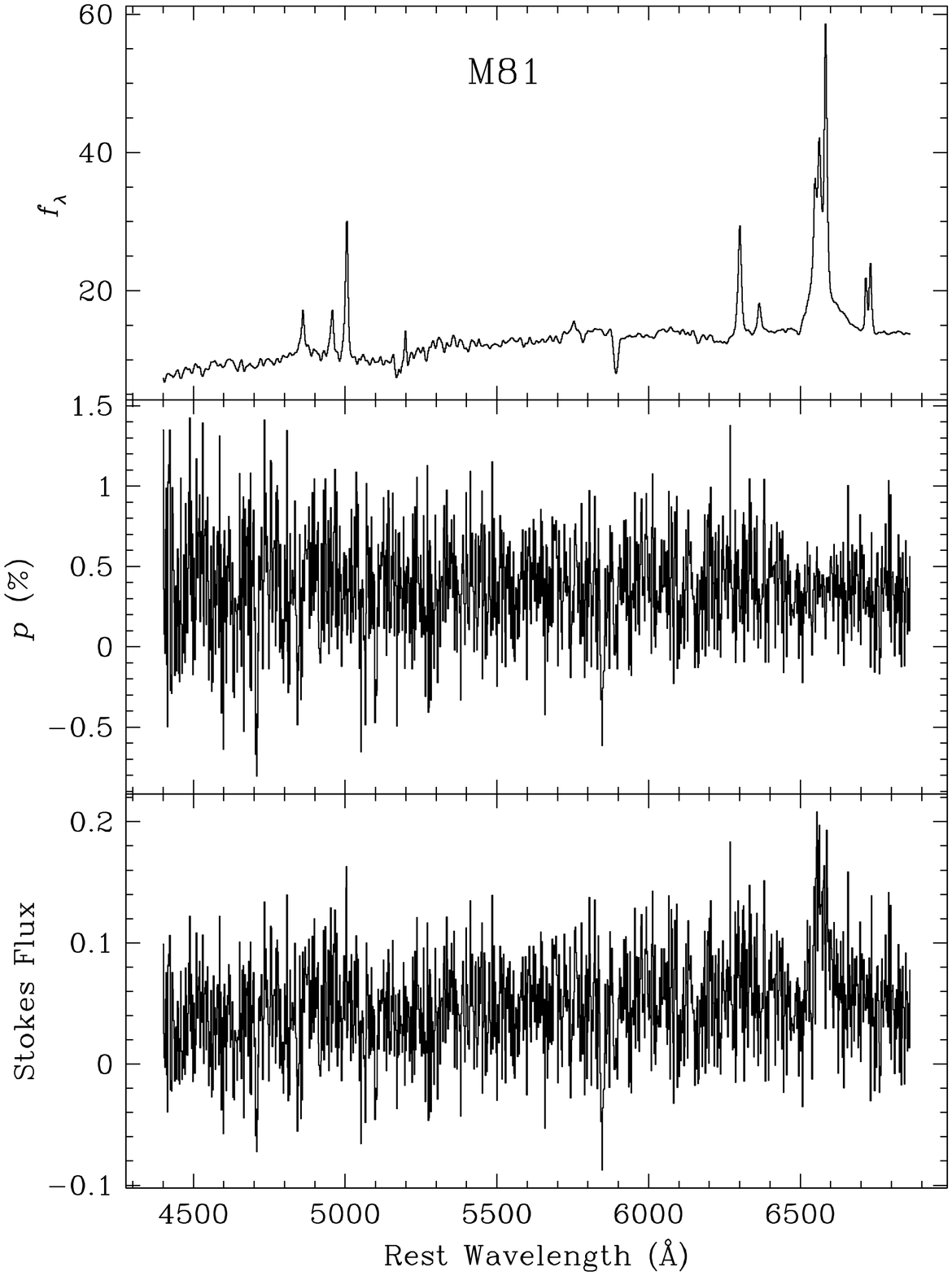]{M81. \label{m81plot}}

\figcaption[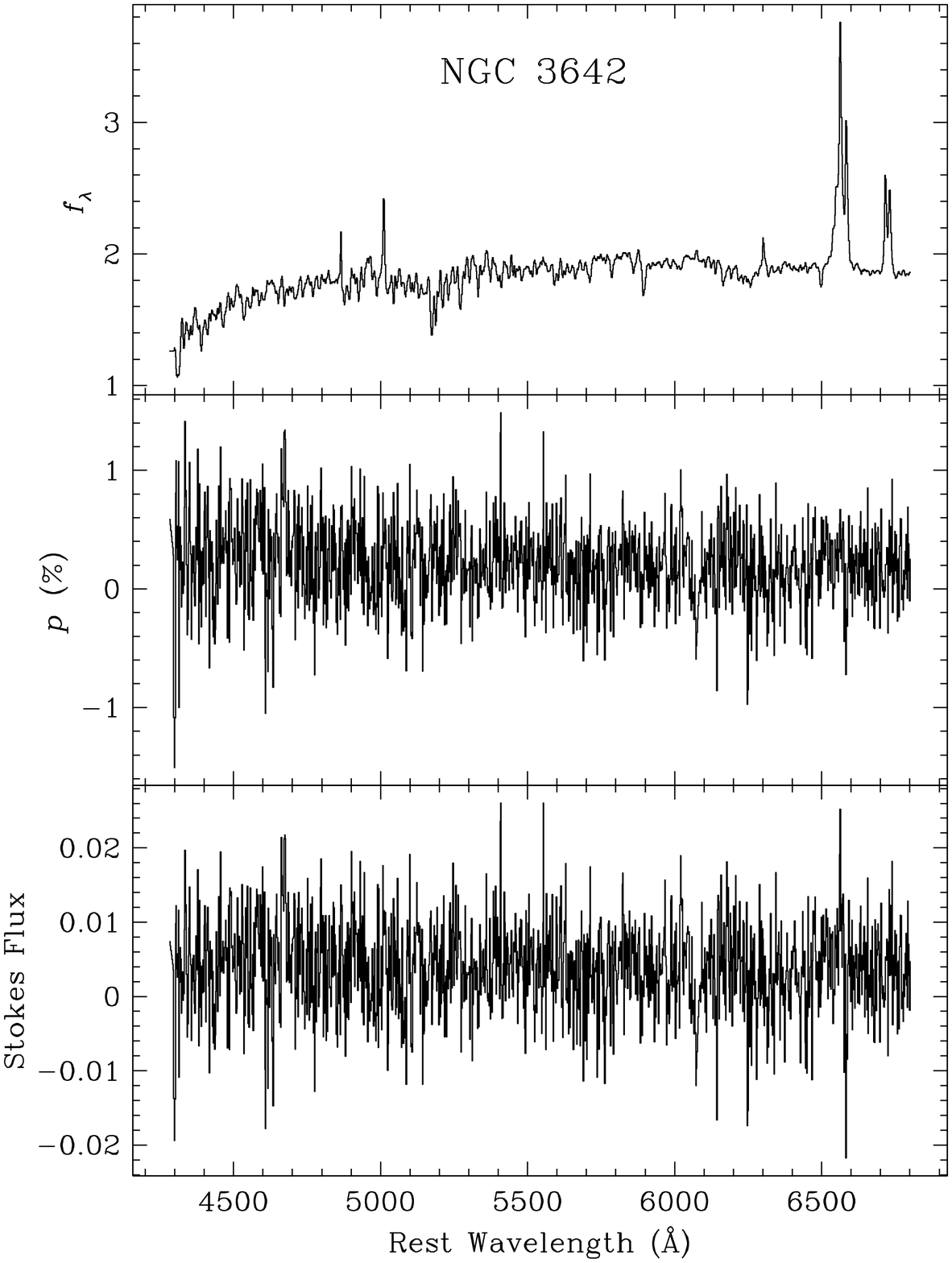]{NGC 3642. \label{ngc3642plot}}

\figcaption[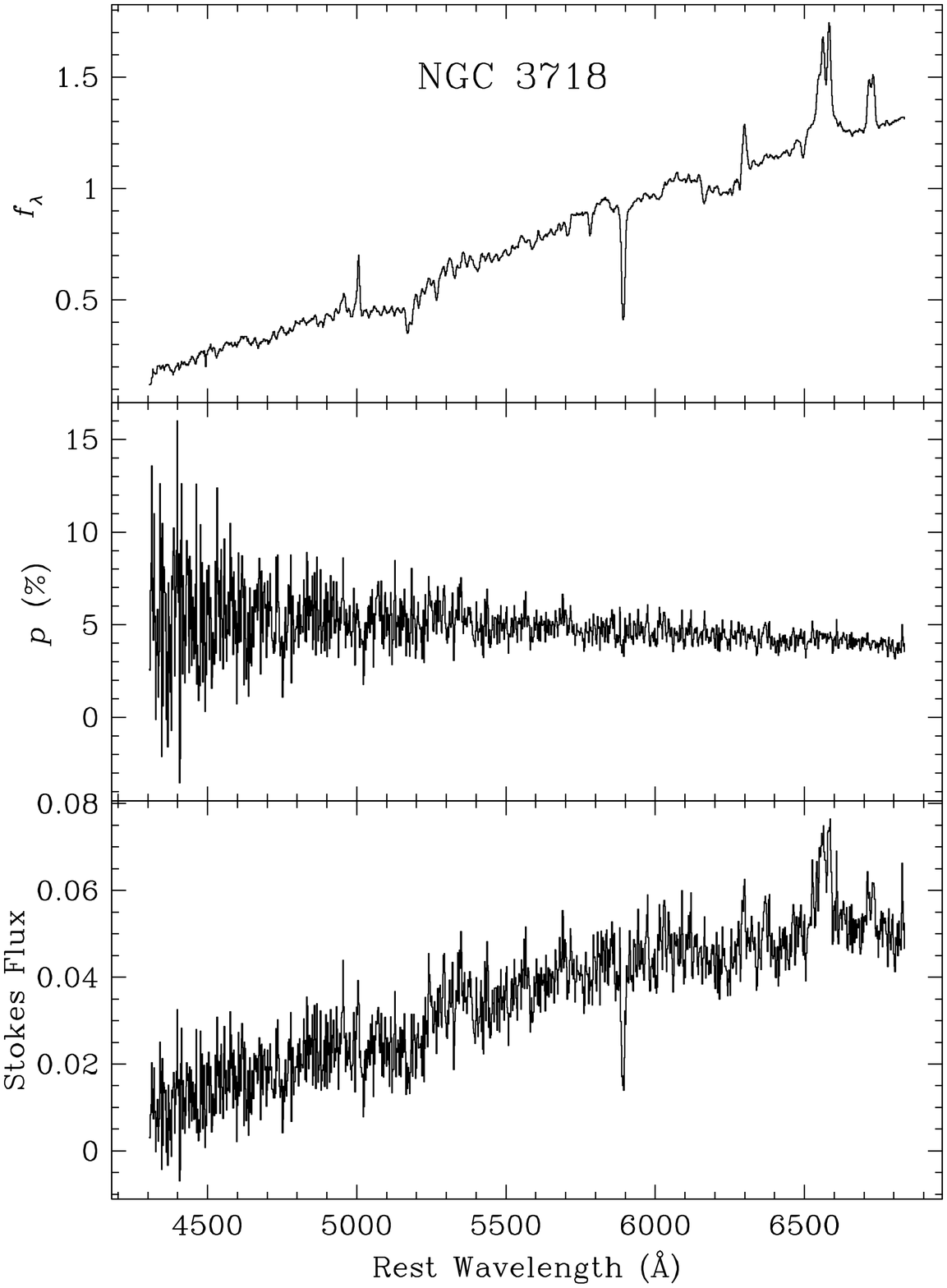]{NGC 3718. \label{ngc3718plot}}

\figcaption[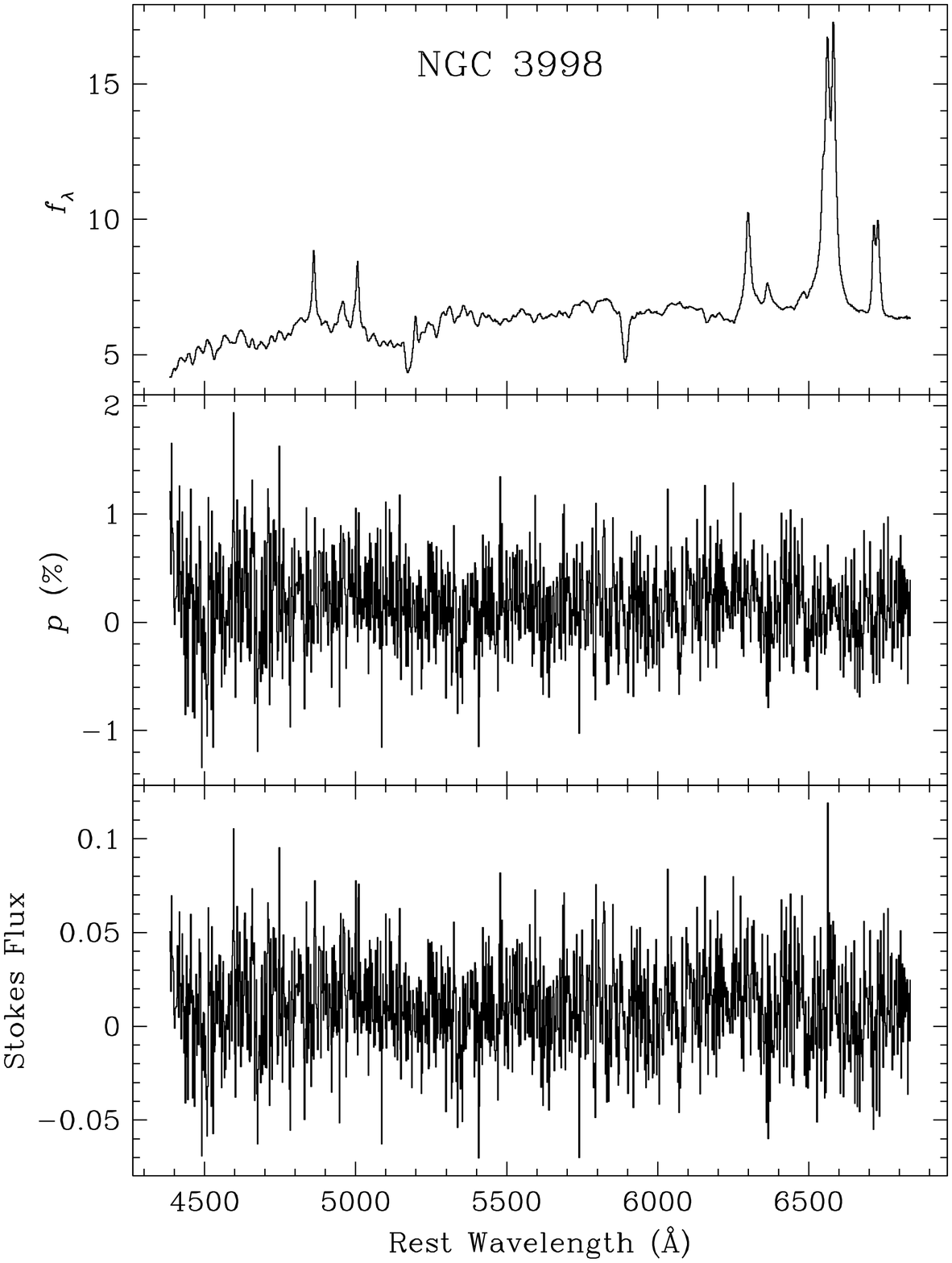]{NGC 3998. \label{ngc3998plot}}

\figcaption[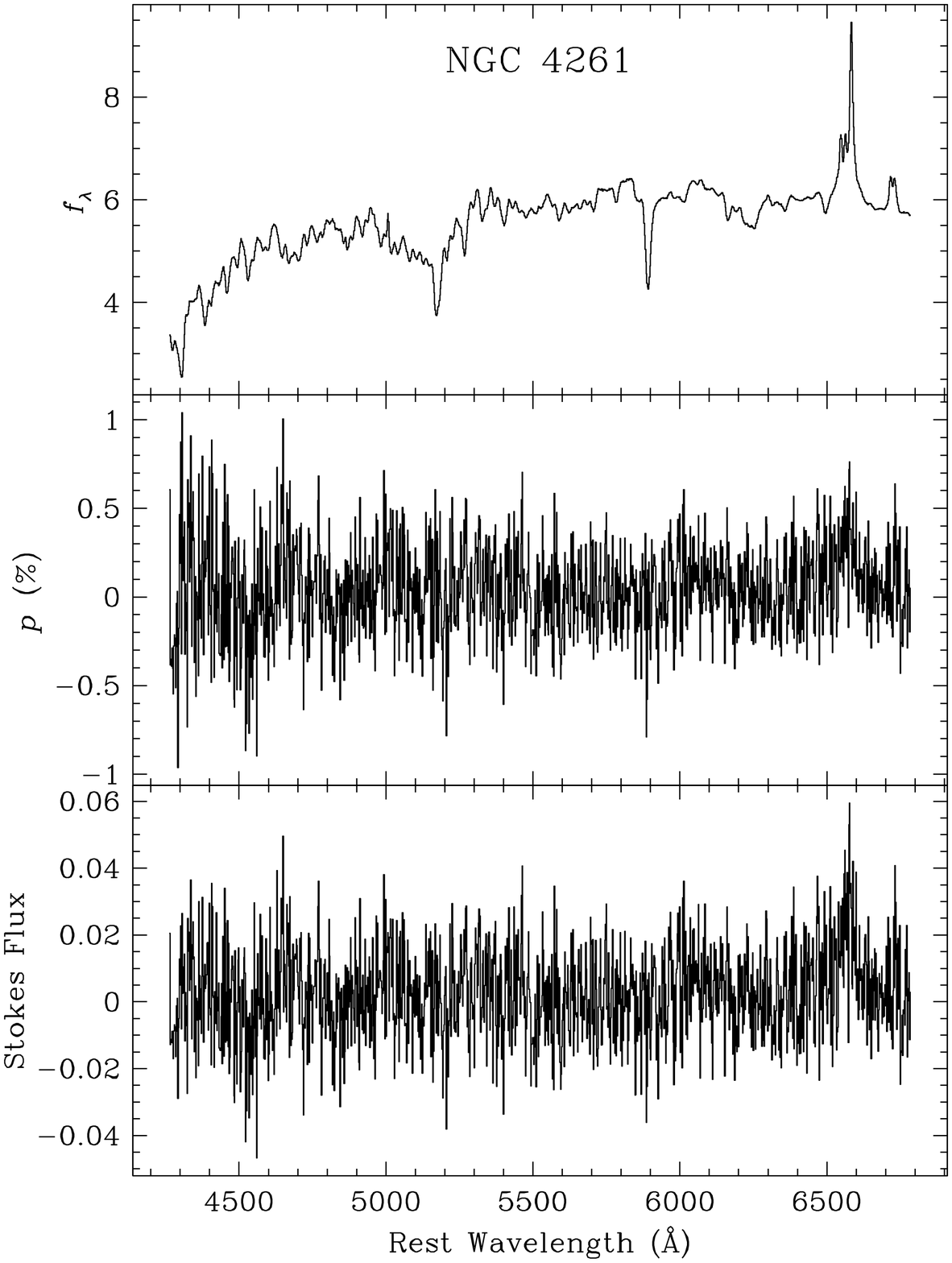]{NGC 4261. \label{ngc4261plot}}

\figcaption[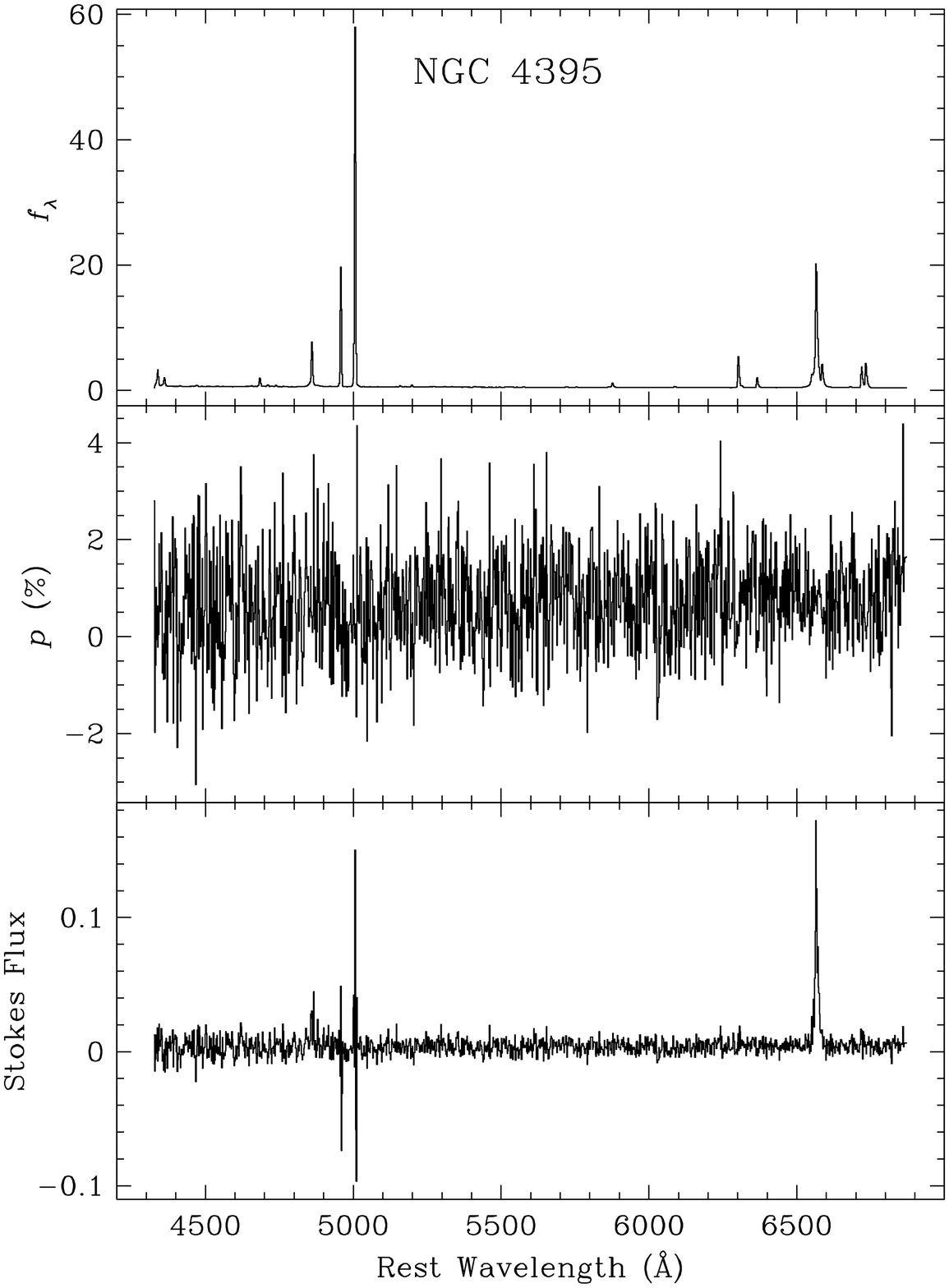]{NGC 4395. \label{ngc4395plot}}

\figcaption[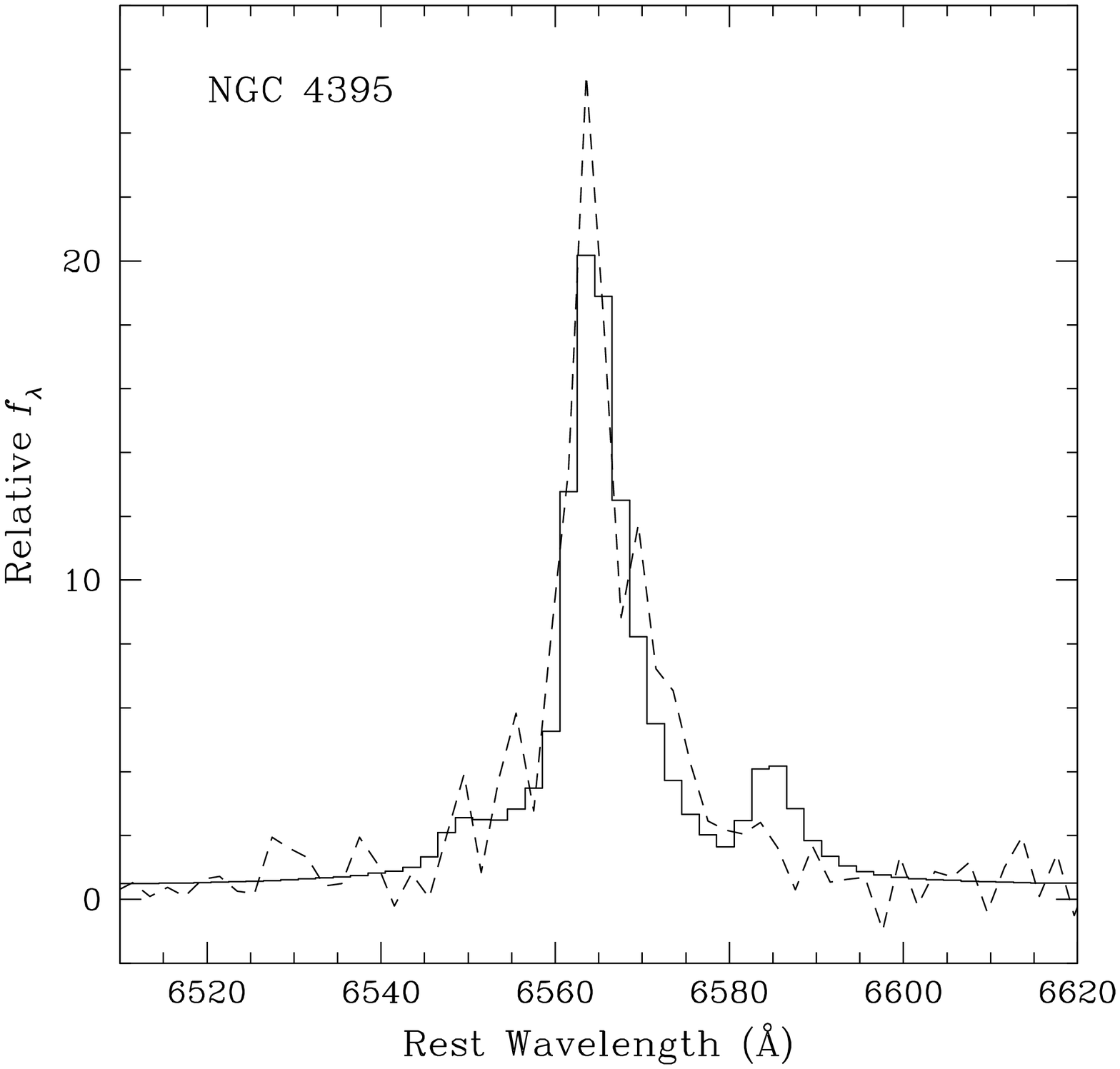]{Comparison of the \hal+[\ion{N}{2}] profiles of
NGC 4395 in total flux and polarized light.  \emph{Solid Line---}
Total flux spectrum.  \emph{Dashed line---} Stokes flux spectrum,
scaled by a factor of 150.  The scaling factor was chosen in order to
match the continua in total and polarized flux, as the continuum
polarization is 0.67\% over 5100-6100 \AA.
\label{ngc4395hal}}

\figcaption[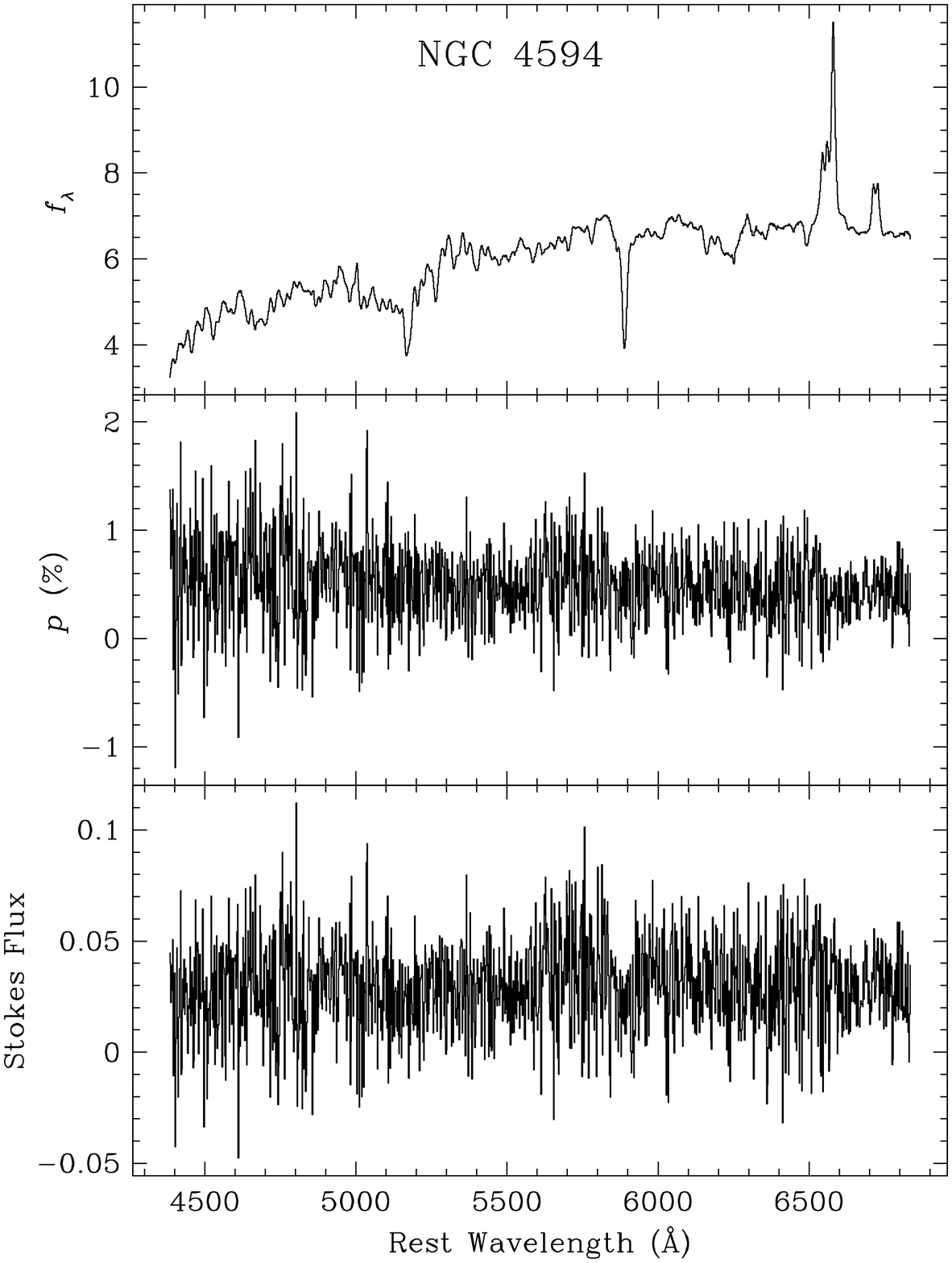]{NGC 4594. \label{ngc4594plot}}

\clearpage

\begin{figure}
\plotone{f1.ps}
\end{figure}

\begin{figure}
\plotone{f2.ps}
\end{figure}

\begin{figure}
\plotone{f3.ps}
\end{figure}

\begin{figure}
\plotone{f4.ps}
\end{figure}

\begin{figure}
\plotone{f5.ps}
\end{figure}

\begin{figure}
\plotone{f6.ps}
\end{figure}

\begin{figure}
\plotone{f7.ps}
\end{figure}

\begin{figure}
\plotone{f8.ps}
\end{figure}

\begin{figure}
\plotone{f9.ps}
\end{figure}

\begin{figure}
\plotone{f10.ps}
\end{figure}

\begin{figure}
\plotone{f11.ps}
\end{figure}

\begin{figure}
\plotone{f12.ps}
\end{figure}

\begin{figure}
\plotone{f13.ps}
\end{figure}

\begin{figure}
\plotone{f14.ps}
\end{figure}

\begin{figure}
\plotone{f15.ps}
\end{figure}

\begin{figure}
\plotone{f16.ps}
\end{figure}

\end{document}